\documentclass[a4paper,10pt]{article}
\pdfoutput=1 

\usepackage{jheppub} 

\usepackage[T1]{fontenc} 
\usepackage{lmodern} 

\usepackage{xcolor}
\usepackage{mathrsfs,wasysym}
\usepackage{booktabs}
\usepackage{slashed}
\usepackage{cancel}
\usepackage{rotating}

\newcommand{\as}{\alpha_s}
\newcommand{\Ord}{\mathcal{O}}
\let\originalleft\left
\let\originalright\right
\renewcommand{\left}{\mathopen{}\mathclose\bgroup\originalleft}
\renewcommand{\right}{\aftergroup\egroup\originalright}
\def\beq{\begin{equation}}  
\def\eeq{\end{equation}}
\def\({\left(}
\def\){\right)}
\def\[{\left[}
\def\]{\right]}

\let\oldsubsection\subsection
\renewcommand\subsection[2][\subsectiontoc]{%
  \def\subsectiontoc{#2}%
  \oldsubsection[#1]{\boldmath #2}%
}

\let\oldsubsubsection\subsubsection
\renewcommand\subsubsection[2][\subsubsectiontoc]{%
  \def\subsubsectiontoc{#2}%
  \oldsubsubsection[#1]{\boldmath #2}%
}

\allowdisplaybreaks

\graphicspath{{images/}}

\arxivnumber{1902.11125}

\title{A new simple PDF parametrization:\\ improved description of the HERA data}

\author[a]{Marco Bonvini}
\affiliation[a]{INFN, Sezione di Roma 1,\\ Piazzale Aldo Moro~5, 00185 Roma, Italy}

\author[b]{and Francesco Giuli}
\affiliation[b]{University of Rome Tor Vergata and INFN, Sezione di Roma 2,\\ Via della Ricerca Scientifica~1, 00133 Roma, Italy}

\preprint{}

\emailAdd{marco.bonvini@roma1.infn.it}

\abstract{%
We introduce a new parametrization for the parton distribution functions (PDFs)
designed to be flexible in the small-$x$ region.
We implement it in the \texttt{xFitter} open-source PDF fitting tool,
and compare it to the default \texttt{xFitter} parametrization,
widely used for many PDF studies, and notably for the HERAPDF determination.
We find that we can describe the combined inclusive HERA~I+II data using NNLO theory
with a significantly higher quality than HERAPDF2.0:
the $\chi^2$ is reduced by more than 60 units,
having used only four more parameters.
Our result highlights a significant parametrization bias in the default
\texttt{xFitter} parametrization at small $x$,
which would lead to even more dramatic effects when used for higher energy colliders,
where the small-$x$ region is more relevant.
We also find that the inclusion of small-$x$ resummation,
that was shown in previous studies to lead to similar improvements in the fit quality,
further reduces the $\chi^2$ by approximately 30 extra units.
}

\begin{document}

\maketitle

\section{Introduction}

The determination of the collinear parton distribution functions (PDFs),
describing the longitudinal momentum fraction $x$ of partons in the proton,
is a fundamental aspect of perturbative QCD phenomenology in presence of initial state hadrons.
While the scale dependence of PDFs can be described in perturbation theory
through the DGLAP evolution equation, their $x$ dependence cannot be computed
with perturbative technologies, because the PDFs are intrinsically long-distance objects.
The current standard methodology for PDF determination
is based on parametrizing the $x$ dependence of the PDFs at a given
``initial'' scale $\mu_0\sim1$~GeV, and then fit such parameters
by comparing a set of data with the corresponding theoretical predictions.
Such theoretical predictions are obtained by first evolving with the DGLAP equation
the PDFs at the scale of the process of a given datapoint,
and then convoluting with the perturbative partonic cross sections
according to the collinear QCD factorization theorem.

The resulting fitted PDFs thus depend on various aspects of the procedure.
In the first place, the accuracy of the theory used to compute the partonic cross sections
(and DGLAP splitting functions) represents the main characterising aspect
of the extracted PDFs. Such accuracy does not only include the perturbative order
at which such objects are computed, but also the scheme, the way heavy quarks are treated,
the choice of the unphysical scales (e.g.\ the renormalization scale), and so on.
Secondly, the dataset considered provides the other major distinctive aspect of the PDFs.
Most PDF sets are based on the precise HERA DIS data, but many other data are available and often used,
both from other DIS experiments and from LHC and Tevatron colliders.
Thirdly, the choice of the parametrization plays an important role.
A good parametrization is one that is able to describe the data with the least possible bias,
unless this is motivated by physical expectations, while keeping at the same time
a sufficiently small number of parameters in order to avoid overfitting.
Finally, there is further dependence on the input parameters (quark masses, strong coupling),
on the fitting methodology (choice of $\chi^2$ definition, minimization methods, uncertainty estimation),
and on various other technical aspects, which however play a minor role.

Various groups are actively involved in extracting PDFs from experimental data.
The mainstream PDF groups, whose PDFs are customarily used in most LHC studies,
are CT~\cite{Dulat:2015mca}, MMHT~\cite{Harland-Lang:2014zoa} and NNPDF~\cite{Ball:2017nwa},
as recommended by the PDF4LHC working group~\cite{Butterworth:2015oua}.
Other well known PDF fitting groups are ABMP~\cite{Alekhin:2017kpj}, JR~\cite{Jimenez-Delgado:2014twa},
CJ~\cite{Accardi:2016qay}.
A somewhat orthogonal collaboration is the \texttt{xFitter} developer's team~\cite{Alekhin:2014irh,Bertone:2017tig}.
Their goal is to provide an open-source tool (\texttt{xFitter}, formerly \texttt{HERAfitter})
for fitting PDFs that is versatile and accessible to anyone:
indeed, \texttt{xFitter} is widely used by both experimental and theory groups.
Its use is often limited to specific studies (namely, not for the production of general-purpose PDF sets),
with the notable exception of HERAPDF~\cite{Abramowicz:2015mha},
that is the official PDF set released by the H1 and ZEUS collaborations
obtained by fitting the HERA collider data.
Also the ATLAS and CMS collaborations make extensive use of \texttt{xFitter} for studies on the impact of
their data in PDF determination, see e.g.~\cite{Aaboud:2016btc,Aaboud:2016zpd,Sirunyan:2017azo,Sirunyan:2017skj}.

The \texttt{xFitter} tool provides various alternative options for many of the distinctive
aspects characterizing PDF sets.
One of these is the choice of the PDF parametrization at the initial scale.
The default parametrization used in \texttt{xFitter}
(which is the most used in \texttt{xFitter} applications, including HERAPDF)
is a very simple one, namely
\beq\label{eq:OldPar}
xf(x,\mu_0^2) = A\,x^B (1-x)^C \Big[1+Dx+Ex^2\Big] -A'\,x^{B'} (1-x)^{C'}.
\eeq
More precisely, the ``negative term'' (the one dependent on primed parameters)
is implemented\footnote{Our comments refer to the latest \texttt{xFitter} release, version 2.0.0 (Frozen Frog).}
only for the gluon PDF, while the quark PDFs only admit a low degree polynomial
multiplying the asymptotic structure $x^B (1-x)^C$ (first term in Eq.~\eqref{eq:OldPar}).
This parametrization is certainly adequate at large $x$,
while it is not very flexible at smaller $x$, where it does not allow
to create structures, since the shape of the PDF is strongly dominated by the asymptotic behaviour $x^B$.
The parametrization used for the gluon PDF is more flexible in the small-$x$ region
thanks to the presence of an additional contribution with its own asymptotic behaviour $x^{B'}$.
Indeed the gluon deserves more attention, as for instance its shape strongly depends on the
perturbative order of the theory used in the fit.
For example, in a next-to-leading order (NLO) fit the data favour a gluon PDF that grows at small $x$,
while in a next-to-next-to-leading order (NNLO) fit (the current state of the art)
the data favour a gluon PDF that starts decreasing at small $x$ after an initial growth.\footnote
{Note that this behaviour of the gluon is likely an artefact of fixed-order perturbation theory,
that is unstable at small $x$ due to the presence of large logarithms of $x$.
The resummation of such logarithms stabilizes the perturbative expansion~\cite
{Ciafaloni:2003rd,Ciafaloni:2007gf, Altarelli:2005ni,Altarelli:2008aj, Thorne:2001nr,White:2006yh},
and the resulting gluon PDF has a very different shape, rising at small $x$~\cite{Ball:2017otu,Abdolmaleki:2018jln}.}
The ``negative term'' gives additional degrees of freedom that can better describe such shapes
and allow for a more reliable determination of the uncertainty.
However, we note that even with that term the flexibility of the parametrization at small $x$ is somewhat limited.

These considerations
motivated us to explore different parametrizations that allow
the PDFs to have a non-trivial structure at intermediate and small $x$.
This is also very important in the light of future higher-energy colliders,
such as the proposed Large Hadron-electron Collider (LHeC) or the Future Circular
electron-hadron or hadron-hadron Colliders (FCC-eh and FCC-hh).
Indeed, at higher energies smaller values of $x$ will become accessible,
and the PDFs can be constrained at small $x$ with higher precision,
calling for an adequate and flexible functional form for the PDF parametrization to be used in the fit.

Before starting, we stress that this work is not intended as a thorough PDF study.
Rather, our goal is to propose a new PDF parametrization and to make a direct comparison
with the default \texttt{xFitter} one, Eq.~\eqref{eq:OldPar}.
We thus focus on a single dataset and on a single perturbative order.
The resulting PDFs are not meant to be general purpose, and we therefore do not make them public,
although they are available from the authors upon request.

The paper is organised as follows.
In Sect.~\ref{sec:newPar} we present our proposal for a new, still simple,
yet more flexible parametrization.
In Sect.~\ref{sec:fitFO} we present results for a fit to the combined inclusive HERA I+II data
and compare them with HERAPDF2.0 and other PDFs on the market.
In particular, we will find a significant reduction of the $\chi^2$ when using our new parametrization
as opposed to the default \texttt{xFitter} one.
In these fits we use fixed-order theory at NNLO.
In Sect.~\ref{sec:fitFONLL} we study the stability of the PDF determination
upon variations of the parametrization and of theoretical settings.
In Sect.~\ref{sec:fitRes} we perform additional fits including the resummation of small-$x$ logarithms,
which is interesting because the effect of the resummation is to change the shape of the PDFs in the small-$x$ region,
that is where our parametrization gives more flexibility.
We conclude in Sect.~\ref{sec:conclusions}.

\section{The new parametrization}
\label{sec:newPar}

In this section we present our new proposal for a flexible simple parametrization
that can be successfully used to determine PDFs.
We have implemented our proposal in the open-source \texttt{xFitter} package.
To make a quantitative comparison within \texttt{xFitter},
we thus consider the default \texttt{xFitter} parametrization
presented in the introduction, Eq.~\eqref{eq:OldPar}.
This parametrization is used to fit the inclusive HERA data in Ref.~\cite{Abramowicz:2015mha},
leading to the so-called HERAPDF2.0 set.
More specifically, the parametrization used in the HERAPDF2.0 set at the initial scale $\mu_0$ is
\begin{subequations}\label{eq:HERApar}
\begin{align}
  xg(x,\mu_0^2) &= A_g\,x^{B_g} (1-x)^{C_g} -A_g'\,x^{B_g'} (1-x)^{C_g'} \label{eq:gluonOldPar}\\
  xu_v(x,\mu_0^2) &= A_{u_v}\,x^{B_{u_v}} (1-x)^{C_{u_v}} \Big[1+E_{u_v}x^2\Big] \\
  xd_v(x,\mu_0^2) &= A_{d_v}\,x^{B_{d_v}} (1-x)^{C_{d_v}} \\
  x\bar u(x,\mu_0^2) &= A_{\bar u}\,x^{B_{\bar u}} (1-x)^{C_{\bar u}} \Big[1+D_{\bar u}x\Big] \\
  x\bar d(x,\mu_0^2) &= A_{\bar d}\,x^{B_{\bar d}} (1-x)^{C_{\bar d}} \\
  xs(x,\mu_0^2) = x\bar s(x,\mu_0^2) &= r_s\, x\bar d(x,\mu_0^2) \qquad r_s=\frac{f_s}{1-f_s} \quad\text{with $f_s=0.4$ fixed,}
\label{eq:sdef}
\end{align}
\end{subequations}
where the choice of the ``flavour basis'' for the parametrization is motivated 
by the fact that the PDFs $u_v=u-\bar u$ and $d_v=d-\bar d$ have a simple ``valence-like'' shape.
The various parameters are not all free: there are further conditions that link them.
One condition is due to the quark-number and momentum sum rules, that provide three constraints
used to fix the normalization of the gluon and the valence quarks, $A_g$, $A_{u_v}$ and $A_{d_v}$
(see Appendix~\ref{sec:app}).
Another condition is related to the behaviour of the sea distributions.
In particular, the $\bar u$ and $\bar d$ distributions
are forced to behave identically at small $x$, thus fixing
\beq\label{eq:barudcond}
A_{\bar u} = A_{\bar d}, \qquad 
B_{\bar u} = B_{\bar d}.
\eeq
The strange quark is taken to be a fixed fraction of the $\bar d$ distribution,
because the inclusive HERA data do not have enough sensibility on the strange alone.\footnote
{Indeed, inclusive DIS in neutral current only depends on the combinations $d+s$ and $\bar d+\bar s$,
  so the constraints on the strange distributions only come from the (few) charged-current data
  and from the scaling violations.}
Note however that the equality Eq.~\eqref{eq:sdef} can only be valid at the initial scale,
because DGLAP evolution breaks that relation at any other scale.
Moreover, the equality $A_{\bar u} = A_{\bar d}$ implicitly depends on the parameter $r_s$
(or equivalently $f_s$), because it is the sum $\bar d+\bar s$ to be constrained by the data.
Therefore, the conditions Eq.~\eqref{eq:sdef} and Eq.~\eqref{eq:barudcond} have to be taken
with a grain of salt, and their validity should be checked explicitly, e.g.\ by relaxing them
and fitting the corresponding parameters, or by varying $r_s$ (both options are explored by HERAPDF2.0).
Finally, the $C_g'$ parameter is not fitted but fixed to a large value, $C_g'=25$,
to ensure that the negative term of the gluon has an impact only at small $x$.
Taking into account all these conditions, the total number of fitted parameters in HERAPDF2.0 is 14.

This number may seem small, given the fact that one is fitting five PDFs whose shape
is not (fully) prescribed by the theory.
However the HERAPDF2.0 study demonstrated that adding more parameters ($Dx$ and $Ex^2$ terms)
to Eq.~\eqref{eq:HERApar} does not improve the description of the data in a significant way.
This argument is however biased by the functional form adopted.
Indeed, adding more parameters simply means adding extra integer positive powers of $x$
in the polynomial, that can give more flexibility in the large $x$ region (roughly, $x\gtrsim0.1$),
but cannot change in a significant way the shape of the PDF for smaller values of $x$.
For this reason, other groups used a polynomial in $\sqrt{x}$~\cite{Harland-Lang:2014zoa,Dulat:2015mca},
that has the power of having an effect on the PDF shape at smaller values of $x$.
This is certainly a better option, even though at some point the PDF behaviour
will still be fully determined by the $x^B$ term.
This may be a limitation for future higher energy experiments, such as the proposed LHeC or FCC,
but also even for the High-Luminosity phase of the LHC, that will provide more precise
data at small $x$ which may require more flexibility in that region.
One could perhaps consider a polynomial in a smaller power of $x$, e.g.\ $x^{1/3}$ or $x^{1/4}$,
to further extend to smaller values of $x$ the sensibility to such contributions.
However, this can be done only at the price of introducing several new parameters.\footnote
{In the literature other functional forms have been explored, with more complicated structures,
e.g.\ using exponentials (see e.g.\ Refs.~\cite{Gao:2013xoa,Alekhin:2017kpj}). Some of them may work better at small $x$.
Making a thorough comparison is beyond the scope of this work.}

Here we propose a simple extension of the PDF parametrization Eq.~\eqref{eq:OldPar}, designed to be
flexible in both the small- and large-$x$ regions, without introducing too many parameters.
The idea is very simple: on top of a polynomial in $x$ (that gives flexibility to the large-$x$ region),
we consider a polynomial in $\log x$ (that gives flexibility to the small-$x$ region).
The two polynomials can be combined in different ways.
We have considered a multiplicative option
\beq\label{eq:NewParMult}
xf(x,\mu_0^2) = A\,x^B (1-x)^C \Big[1+Dx+Ex^2\Big] \Big[1+F\log x+G\log^2x+H\log^3x\Big]
\eeq
and an additive option
\beq\label{eq:NewParAdd}
xf(x,\mu_0^2) = A\,x^B (1-x)^C \Big[1+Dx+Ex^2+F\log x+G\log^2x+H\log^3x\Big],
\eeq
and we have kept the degree two of the $x$ polynomial, while we have chosen to
reach degree three for the $\log x$ polynomial (this will be motivated later).
Since $\log x\to 0$ as $x\to1$, the logarithmic part of the parametrization has
no impact at large $x$, where the $x$ polynomial is supposed to model the shape.
Similarly, at small $x$ the polynomial in $\log x$ provides the desired shaping effect
that cannot be achieved using a polynomial in $x$.

The difference between the multiplicative and additive options resides in a region
of intermediate $x$'s, roughly around $x\sim0.1$, where
the contributions of the form $x^a\log^bx$ with $a,b>0$ produced in the multiplicative
case but not present in the additive case give a sizeable effect.
Such cross-product terms may in principle give additional flexibility in that region;
however, their coefficients are not independent from the others, de facto leading to more rigidity.
In the additive case, instead, there is a more net separation:
the $F$, $G$ and $H$ coefficients are determined from the small-$x$ region
(and also have an effect at intermediate $x$),
and the $D$ and $E$ coefficients adjust the shape at intermediate and large $x$.
These considerations suggest that the additive option, Eq.~\eqref{eq:NewParAdd},
is advisable.
In our checks we have indeed verified that the multiplicative parametrization
generates unexpected shapes (e.g., small bumps),
while the additive parametrization leads to smoother shapes and to smaller $\chi^2$ in the fit.
We therefore discard Eq.~\eqref{eq:NewParMult} and use Eq.~\eqref{eq:NewParAdd}
throughout the paper.

The polynomial in $\log x$ is taken to be of degree three.
The main motivation for this is the modeling of the gluon PDF.
We have already commented that the gluon PDF determined using NNLO theory has a peculiar shape
at small $x$.
In order to increase the chances of being able to reproduce such a shape accurately
we have decided to also include a $\log^3x$ term.
As far as HERA data are concerned, we will see that we are actually able to describe the gluon
with just the first two powers of $\log x$; however, at future colliders
the highest precision on the small-$x$ gluon may require the inclusion of the $\log^3x$ term.
We stress that this way of describing the gluon PDF is superior with respect to the
\texttt{xFitter} parametrization Eq.~\eqref{eq:gluonOldPar}, because it allows
to produce small-$x$ structures with more flexibility: for instance, Eq.~\eqref{eq:gluonOldPar}
can only have a maximum (and consequently a decreasing asymptotic behaviour at small-$x$),
while our parametrization Eq.~\eqref{eq:NewParAdd} can produce as many stationary points as the degree of the $\log x$ polynomial,
and can also for instance asymptotically go to zero if $B_g>0$\footnote
{Note that these considerations do not consider the physical expectations.
  Regge theory suggests that the gluon PDF should rise at small $x$ as a power, namely $B_g<0$.}
(the last is true also for the HERAPDF2.0 parametrization, provided both $B_g$ and $B_g'$ are positive).

We can now present the actual parametrization used in our fits to the inclusive HERA data.
Our choice is the result of a number of tests, where we started from
a parametrization with a large number of parameters and progressively turned them off on
the basis of their significance (estimated from the relative uncertainty on the parameter).
The minimal parametrization we ended up with is
\begin{subequations}\label{eq:NewPar}
\begin{align}
  xg(x,\mu_0^2) &= A_g\,x^{B_g} (1-x)^{C_g} \Big[1+F_g\log x+G_g\log^2x\Big] \\
  xu_v(x,\mu_0^2) &= A_{u_v}\,x^{B_{u_v}} (1-x)^{C_{u_v}} \Big[1+E_{u_v}x^2+F_{u_v}\log x+G_{u_v}\log^2x\Big] \\
  xd_v(x,\mu_0^2) &= A_{d_v}\,x^{B_{d_v}} (1-x)^{C_{d_v}} \\
  x\bar u(x,\mu_0^2) &= A_{\bar u}\,x^{B_{\bar u}} (1-x)^{C_{\bar u}} \Big[1+D_{\bar u}x+F_{\bar u}\log x\Big] \\
  x\bar d(x,\mu_0^2) &= A_{\bar d}\,x^{B_{\bar d}} (1-x)^{C_{\bar d}} \Big[1+D_{\bar d}x+F_{\bar d}\log x\Big],
\end{align}
\end{subequations}
that leads to the best $\chi^2/\text{dof}$ for fixed-order NNLO fit.
In particular, turning back on any additional parameter to this minimal parametrization
the $\chi^2$ either remains the same or decreases by at most one unit.
We keep using the same conditions of the HERAPDF2.0 set.
Namely, we take the strange to be a fixed fraction of the $\bar d$
distribution, and we assume that the small-$x$ behaviour of $\bar u$ and $\bar d$ is the same.
We implemented this last condition by also fixing the coefficients of the logarithmic terms
to be the same, namely $F_{\bar u}=F_{\bar d}$
(the same condition would also apply to the $G$ and $H$ terms, if added).
Of course, sum rules are used to fix the normalization of the gluon and the valence quarks.
We stress that the Mellin transform of the PDFs can be easily computed analytically for this parametrization,
to provide the best numerical performance for the implementation of the sum rules
(see Appendix~\ref{sec:app} for further detail).

Our proposed parametrization Eq.~\eqref{eq:NewPar} depends on 18 free parameters that must be fitted.
This is to be compared with the HERAPDF2.0 parametrization, Eq.~\eqref{eq:HERApar},
that depends on 14 free parameters.
In particular, our parametrization has two extra parameters for the $u_v$, and two for $\bar u$ and $\bar d$.
Despite the small number of extra parameters (only four), we will see in the next section
that we achieve a reduction of the $\chi^2$ of more than 60 units with respect to HERAPDF2.0.
Most notably, the major improvement comes from the gluon PDF, where the number of free parameters
is the same.

The new parametrization also offers new handles for estimating the ``parametrization uncertainty'',
namely the bias induced by the choice of the parametrization.
One way to do so is by turning on (or off) some of the parameters and compare the resulting PDFs
with the default ones.
After playing with the parameters, we ended up with the following list of parameters
giving the most significant changes in the PDFs while leaving the $\chi^2$ almost unchanged:
\beq
F_{d_v}, \quad
D_g, \quad
H_g(F_g=0).
\eeq
The last item of the list means that the linear logarithmic term $F_g$
is deactivated and the cubic logarithmic term $H_g$ is activated at the same time.
The reason for this choice is that simply activating $H_g$ does not give any effect,
as its value determined by the fit is compatible with zero.
However, trading the linear term with a cubic term does change the PDF,
while giving a description of the data of the same quality.
Other parameters do not give appreciable effects and are thus not used.

Before moving to the PDF fits, we want to comment on the physical adequacy of the
functional form adopted in our parametrization.
Despite a logarithm is subleading with respect to a power, the presence of the $\log^k x$ terms
changes the asymptotic small-$x$ behaviour of the PDFs, making it richer than just a $x^B$ behaviour.
This may seem to contradict some expectations based on theoretical considerations (Regge theory),
that predict a power-like behaviour at small $x$.
There are however a number of considerations to take into account.
\begin{itemize}
\item A power-like behaviour is a prediction of Regge theory~\cite{Collins:1977jy}.
  However, Regge theory is only a leading description of the small-$x$ region, and it is well likely that subleading
  contributions can correct such behaviour with logarithmic contributions.
\item Even if a power-like behaviour could be expected, this can only happen at a given scale, 
  since perturbative DGLAP evolution induces logarithmic dependence on the PDFs at small $x$ at any other scale
  through the logarithms present in the splitting functions.
  So a power-like parametrization could be appropriate at a single scale at most,
  and at any other scale a functional form of the PDFs that includes logarithms is not only allowed, but expected.
  Since it is not known at which scale a pure power-like behaviour should be expected (if any),
  our proposed functional form is perfectly legitimate.
\item We also recall that we are fitting in a finite region of $x$, so we are not really reaching the asymptotic behaviour.
  This means that our parametrization may not be appropriate at very small $x$,
  while performing well in the region of $x$ accessible by the HERA data ($x\gtrsim5\times10^{-5}$).
\end{itemize}
So, in conclusion, it seems very fair to consider our parametrization as perfectly valid,
both from a theoretical and from a practical point of view.

\section{Fixed-order PDF determination and comparison with HERAPDF2.0}
\label{sec:fitFO}

We are now ready to present the results of the PDF fits.
In this section we focus on the fixed-order, that we take to be NNLO (the highest order available today).
In order to directly compare with HERAPDF2.0, we use the same definition of the $\chi^2$, namely~\cite{Abramowicz:2015mha}
\beq\label{eq:chi2}
  \chi^2 = \sum_i\frac{\[ D_i -T_i\(1-\sum_j\gamma_{ij} b_j\) \]^2}
  {\delta_{i,{\rm uncor}}^2 T_i^2+\delta_{i,{\rm stat}}^2D_i T_i}
  + \sum_j b_j^2
  + \sum_i \log \frac{\delta_{i,{\rm uncor}}^2 T_i^2+\delta_{i,{\rm stat}}^2D_i T_i}
  {\delta_{i,{\rm uncor}}^2 D_i^2+\delta_{i,{\rm stat}}^2D_i^2},
\eeq
where $D_i$ represent the measured data, $T_i$ the corresponding theoretical prediction,
$\delta_{i,\rm uncor}$ and $\delta_{i,\rm stat}$ are the uncorrelated systematic and the statistical uncertainties
on $D_i$, and correlated systematics are described by $\gamma_{ij}$ and are accounted for
using the nuisance parameters $b_j$.
The sums over $i$ extend over all data points, while the sum over $j$ runs over the various sources
of correlated systematics.

As we commented in the introduction, there are many technical aspects which a PDF fit depends upon.
One of these is the scheme used to deal with heavy quarks.
The scheme used in HERAPDF2.0 is the ``optimized'' version~\cite{Thorne:2012az}
of the Thorne-Roberts (TR) scheme~\cite{Thorne:1997ga,Thorne:2006qt}, that gives the best performance in describing the data at NNLO.
Commenting on this choice is beyond the scope of this work,
but we have stressed it because we will consider a different scheme in the next Sect.~\ref{sec:fitFONLL}.
In this section, we use exactly the same setting of HERAPDF2.0,
including quark masses, initial scale, etc., in order to get exactly the same
PDFs if using the HERAPDF2.0 parametrization Eq.~\eqref{eq:HERApar}.

\begin{table}
\centering
\begin{tabular}{lcc}
  Contribution to $\chi^2$ &HERAPDF2.0 &  Our fit (new parametrization)  \\
  \midrule
  subset NC $e^+$ 920    $\tilde\chi^2/\rm{n.d.p.}$   & $444/377$   & $403/377$   \\
  subset NC $e^+$ 820    $\tilde\chi^2/\rm{n.d.p.}$   & $ 66/ 70$   & $ 74/ 70$   \\
  subset NC $e^+$ 575    $\tilde\chi^2/\rm{n.d.p.}$   & $219/254$   & $221/254$   \\
  subset NC $e^+$ 460    $\tilde\chi^2/\rm{n.d.p.}$   & $217/204$   & $222/204$   \\
  subset NC $e^-$        $\tilde\chi^2/\rm{n.d.p.}$   & $219/159$   & $220/159$   \\
  subset CC $e^+$        $\tilde\chi^2/\rm{n.d.p.}$   & $ 45/ 39$   & $ 38/ 39$   \\
  subset CC $e^-$        $\tilde\chi^2/\rm{n.d.p.}$   & $ 56/ 42$   & $ 50/ 42$   \\
  correlation term + log term     & $91 + 5$ & $75 - 3$ \\
  \bf\boldmath Total $\chi^2/\rm{d.o.f.}$  &\boldmath $1363/1131$ &\boldmath $1301/1127$ \\
\end{tabular}
\caption{Total $\chi^2$ per degrees of freedom (d.o.f.)
  and the partial $\tilde\chi^2$ (first term of Eq.~\eqref{eq:chi2})
  per number of data points (n.d.p.) of each subset of the inclusive HERA dataset,
  for HERAPDF2.0 and our fit obtained with the parametrization Eq.~\eqref{eq:NewPar}.
  The second and third terms of Eq.~\eqref{eq:chi2}, denoted correlation and log terms respectively,
  are also shown.}
\label{tab:chi2TR}
\end{table}

We begin by presenting the results of the fit in terms of $\chi^2$,
comparing with HERAPDF2.0, i.e.\ the very same fit but with the default \texttt{xFitter} parametrization.
The numbers are given in Tab.~\ref{tab:chi2TR}, where on top of the total $\chi^2$,
also the individual contributions
from each subset composing the combined HERA I+II inclusive dataset
to the first (``$\tilde\chi^2$'') term of the $\chi^2$ Eq.~\eqref{eq:chi2} are shown,
as well as the total second (``correlation'') and third (``log'') terms of Eq.~\eqref{eq:chi2}.

We observe a dramatic reduction of the total $\chi^2$ when using our new parametrization:
from 1363 to 1301. This reduction of 62 units of $\chi^2$ is much larger than the increase
of 4 units in the number of parameters.
Indeed, the $\chi^2$ divided by the number of degrees of freedom reduces from
1.21 to 1.15, which is a significant improvement for 1145 datapoints.
This reduction is mostly due to a better description of the neutral-current $E_p=920$~GeV dataset,
which improves by 41 units.
This dataset contains the datapoints at smaller $x$, that are indeed responsible for most
of the improvement, as we will see later.
The other significant reduction is in the correlation term, reducing from 91 to 75.
This implies that the theoretical prediction agrees better with the data without the need
of using a lot of correlated systematic shifts (the sum in the numerator of the first term of Eq.~\eqref{eq:chi2}).
For the other datasets the variation is milder.

This dramatic reduction of $\chi^2$ highlights a significant bias in the form of the parametrization
adopted in HERAPDF2.0, namely the default \texttt{xFitter} parametrization, Eq.~\eqref{eq:HERApar}.
We have already argued this on the basis of mathematical considerations,
and we see it in practice from the result of the fit.
In other words, the shape of the PDFs favoured by HERA data according to NNLO theory
cannot be accurately described by the parametrization Eq.~\eqref{eq:OldPar}.
Instead, our newly proposed parametrization Eq.~\eqref{eq:NewParAdd}
gives much better performances.

It is difficult to quantify exactly how our result compares with other parametrizations used
by other PDF fitting groups without performing a fit ourself with identical settings.
However, such a task is not trivial, and it is beyond the scope of this paper.
In order to give an idea, we simply report here the values of the $\chi^2$
of the combined inclusive HERA I+II dataset reported by the mainstream PDF collaborations.
The MMHT collaboration reports $\chi^2=1319$~\cite{Harland-Lang:2016yfn},
while the NNPDF3.1 set gives $\chi^2=1328$~\cite{Ball:2017nwa}.
In both cases, the result refers to the very same dataset with 1145 datapoints that we use.
The CT14 study finds $\chi^2=1402$~\cite{Hou:2016nqm},
however in this case the dataset is smaller, with 1120 datapoints.
The ABMP work reports $\chi^2=1510$~\cite{Alekhin:2017kpj},
with a larger dataset including 1168 datapoints.
We stress that all these $\chi^2$ values are computed from a global PDF fit;
fitting only the HERA dataset would likely give a smaller $\chi^2$.
It has also to be noticed that there are various theoretical aspects
that are different between the sets, so we cannot consider these numbers as
a comparison among parametrizations.
With all these caveats in mind, it is anyway interesting to observe
that our result is never worse than these.
This is encouraging, and it suggests that our parametrization is at the very least competitive
with the ones used in mainstream PDFs.

\begin{figure}[t]
  \centering
  \includegraphics[width=0.328\textwidth,page=1]{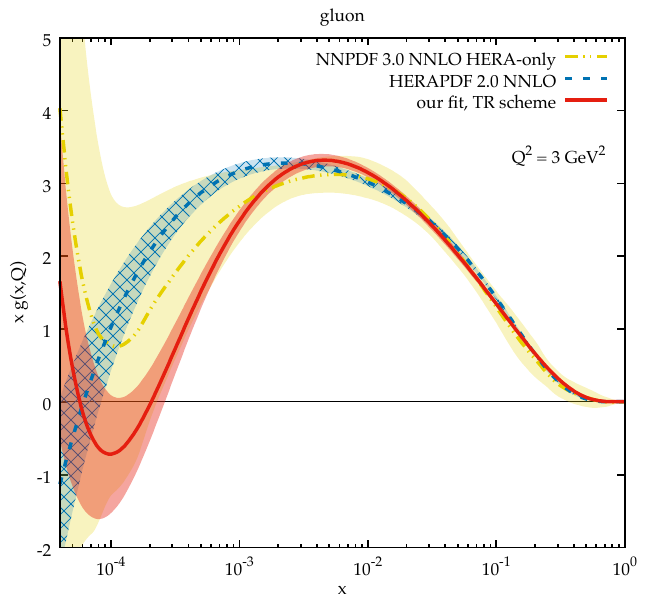}
  \includegraphics[width=0.328\textwidth,page=3]{NNPDF_HERAPDF_comparison.pdf}
  \includegraphics[width=0.328\textwidth,page=5]{NNPDF_HERAPDF_comparison.pdf}\\
  \includegraphics[width=0.328\textwidth,page=2]{NNPDF_HERAPDF_comparison.pdf}
  \includegraphics[width=0.328\textwidth,page=4]{NNPDF_HERAPDF_comparison.pdf}
  \includegraphics[width=0.328\textwidth,page=6]{NNPDF_HERAPDF_comparison.pdf}
  \caption{Comparison of our fit (solid red) with HERAPDF2.0 (dashed blue) and NNPDF3.0 HERA-only (dot-dot-dashed yellow)
    for the gluon, total singlet, $\bar u$, $\bar d+\bar s$, $u_v$ and $d_v$ PDFs.
    The uncertainty shown is only the ``experimental'' one, namely the one coming from
    the uncertainty on the parameters determined from the fit.
    For NNPDF, this uncertainty (typically larger) actually covers other kinds of uncertainties,
    such as those coming from parametrization bias.}
  \label{fig:TR}
\end{figure}

We now move to the comparison of the PDFs. In Fig.~\ref{fig:TR}
we show some representative PDFs\footnote
{We plot the combination $\bar d+\bar s$ because in neutral-current DIS the two PDFs always contribute in this combination,
  as we have already commented in Sect.~\ref{sec:newPar}.}
at the scale $Q^2=3$~GeV$^2$ from HERAPDF2.0 and our new fit.
The PDFs are, of course, qualitatively similar.
However, the shape is in general smoother for HERAPDF2.0
(due to the simpler parametrization),
while our PDFs present a richer structure in the medium-small $x$ region.
The shape of the gluon and of the sea distributions show the largest differences.
In particular, our gluon decreases more rapidly for $x$ below $10^{-2}$,
and then starts rising again for $x<10^{-4}$.
This peculiar form is induced by the asymptotic behaviour dominated by $x^{B_g}G_g\log^2x$ with $B_g<0$ and $G_g>0$
that drives the asymptotic growth
(some comments on this asymptotic behaviour are given in Appendix~\ref{sec:locmin}).
Conversely, the HERAPDF2.0 gluon keeps decreasing due to the dominance of the $-A_g'x^{B_g'}$ term with $A_g'>0$.
Also the up-valence distribution is rather different,
while the down-valence is basically identical,
which is a consequence of the fact that for this PDF the parametrization we use
is identical to HERAPDF2.0.
We observe that despite the presence of the logarithmic terms
the small-$x$ behaviour of all the PDFs except the gluon 
coincides between the two sets. This is probably due to the data constraining
the quark PDFs at small $x\sim10^{-4}\div10^{-3}$.
We also note that in many regions the HERAPDF2.0 uncertainty
(which is only the ``experimental'' one, namely the one obtained from the fit using a $\Delta\chi^2=1$ criterium)
is smaller than ours, which is a consequence of the limited flexibility of the parametrization
Eq.~\eqref{eq:HERApar} at medium/small $x$.

The fact that with the new parametrization the $\chi^2$ has improved significantly
suggests that the PDF shape that we find is favoured by the data.
A question arises naturally: how does it compare with other determinations on the market?
Rather than performing a thorough comparison, we consider
a single alternative PDF determination, from the NNPDF collaboration.
We made this choice because NNPDF has the most flexible parametrization
ever used in PDF determination, with 39 parameters for \emph{each} fitted PDF,
which is the least biased PDF parametrization used in modern PDF sets.
Moreover, we use a (NNPDF3.0) set that has been obtained fitting only HERA data~\cite{Ball:2014uwa},
to make the comparison as fair as possible.\footnote
{We stress that this NNPDF set is based on a dataset including the inclusive HERA data and also charm production data.
The presence of the charm dataset can generate a difference, especially on the gluon PDF that is the most
sensitive to this process. Many other theoretical details are different, e.g.\ the heavy quark scheme adopted.}
This comparison is shown in the same Fig.~\ref{fig:TR}.
Despite the fact that the NNPDF uncertainty is rather large,
we see that in various cases the HERAPDF2.0 PDFs lie outside the NNPDF band,
while our PDFs lie inside it (or at the edge of it).
We also observe that the shape of the NNPDF gluon is very similar to ours,\footnote
{We stress, however, that the central PDF of the NNPDF set is the average of many PDF members each with its own shape,
and as such it does not necessarily correspond to any of such fitted members.
Nevertheless, it is clear that several members do grow at small $x$, otherwise the central PDF
would not have that shape. Therefore, even if comparing
directly with the central NNPDF PDF is not very meaningful,
it is undoubtable that NNPDF finds that the gluon has a rising shape at small $x$ as in our result.}
in particular the rising asymptotic behaviour at small $x$,
which is instead very different from HERAPDF2.0.

\begin{figure}[t]
  \centering
  \includegraphics[width=0.95\textwidth]{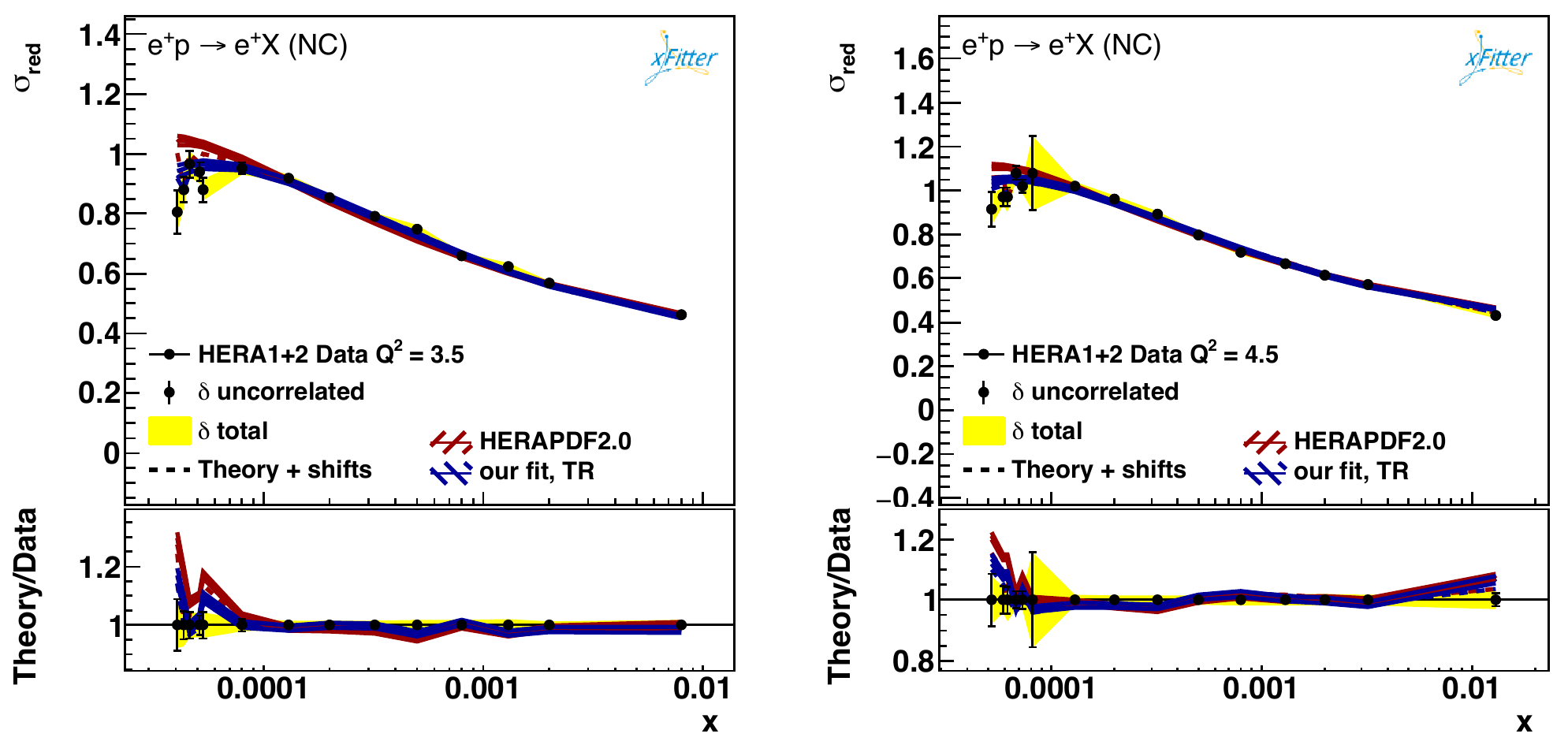}
  \caption{An example of comparison of the theoretical predictions from our fit
    and HERAPDF2.0 with data: low-$Q^2$ neutral-current reduced cross section from the neutral-current $E_p=920$~GeV dataset.}
  \label{fig:datacomp}
\end{figure}

In order to understand how the quality of the fit could improve so much,
we have inspected in detail the comparison of HERA data with the theoretical predictions
using both our fit and HERAPDF2.0. In most of the cases the agreement is at the same level,
in some cases some datapoints are better described by our PDF set, without a precise pattern.
The only exception is the low-$Q^2$ low-$x$ data, where a clear improvement of the theoretical
description is manifest.
The lowest two $Q^2$ bins of our dataset, at $Q^2=3.5$~GeV$^2$ and $Q^2=4.5$~GeV$^2$,
are shown in Fig.~\ref{fig:datacomp}, where the difference between the two descriptions is apparent.
We observe that this region (low-$Q^2$ and low-$x$) is the same where the
impact of the resummation of $\log(1/x)$ terms is expected (and found~\cite{Ball:2017otu,Abdolmaleki:2018jln}) to be largest.
We conclude that \emph{part} of the improvement in the description
of the HERA data comes from the ability of the new parametrization
of being flexible enough at small $x$ to better describe this region.
We also observe that the $\chi^2$ reduction obtained using our parametrization
is of the same size as that obtained with the inclusion of small-$x$ resummation in the theory~\cite{Ball:2017otu,Abdolmaleki:2018jln}.
In order to better understand the interplay of our new parametrization and
the inclusion of small-$x$ resummation, we will also present resummed fits
in Sect.~\ref{sec:fitRes}.

In the next, we will study variations of parametrization and of the initial scale to verify the robustness
of our parametrization, and we will subsequently consider the effect of small-$x$ resummation.
Meantime, we can already fairly conclude that the new parametrization that we propose, Eq.~\eqref{eq:NewParAdd},
gives much more flexibility with respect to the default \texttt{xFitter} parametrization Eq.~\eqref{eq:OldPar},
allowing for a significantly improved description of the HERA data.
This is remarkable given the simplicity of our proposal, and the small number of additional parameters used.
We therefore consider our proposal as a very simple, yet very useful improvement,
and suggest its adoption as a default parametrization in \texttt{xFitter}.

\section{Stability of the fit with the new parametrization}
\label{sec:fitFONLL}

In order to understand the robustness of our parametrization,
we now study the impact of several possible variations related to it.
The simplest one is the variation of the scale $\mu_0$ at which the PDFs are parametrized.
If the parametrization is flexible enough, it should be able to accomodate the
small amount of DGLAP evolution between two choices of initial scales
that are not too far away from each other.
Any strong dependence on the choice of $\mu_0$ highlights a parametrization bias.
Additional variations include the possibility of adding (or removing) some parameters
to the PDF parametrization, as already discussed in Sect.~\ref{sec:newPar}.
Other parameters such as the strong coupling or the quark masses are not directly
connected to the parametrization, and therefore we do not consider their variation in this study.

We recall that we are also interested in performing fits including the resummation of small-$x$ logarithms.
This is easily achieved using the \texttt{APFEL} code~\cite{Bertone:2013vaa},
since it has been interfaced to the \texttt{HELL} code~\cite{Bonvini:2016wki,Bonvini:2017ogt,Bonvini:2018xvt,Bonvini:2018iwt}
that provides the resummed coefficient functions and splitting functions.
The \texttt{APFEL+HELL} bundle is available in \texttt{xFitter}~\cite{Abdolmaleki:2018jln},
however \texttt{APFEL} does not implement the TR scheme~\cite{Thorne:2012az,Thorne:1997ga,Thorne:2006qt},
but rather the FONLL scheme~\cite{Forte:2010ta}.
This forces us to use FONLL for resummed fits,\footnote
{In principle, it should be straightforward to implement the resummed contributions in the TR
  or any other scheme, thanks to the results presented in Ref.~\cite{Bonvini:2017ogt}
  and the equivalence among the schemes investigated in Refs.~\cite{Bonvini:2015pxa,Ball:2015dpa}.
  In practice, however, this requires modifying existing codes.
  For instance, the TR scheme structure functions are constructed within \texttt{xFitter}
  out of the massless and massive structure functions provided by the \texttt{QCDNUM} package~\cite{Botje:2010ay}.
  Adding resummation to the TR scheme coefficient functions would then require a delicate
  work of interfacing \texttt{QCDNUM} with \texttt{HELL} and a careful validation,
  similarly to what has been done for \texttt{APFEL+HELL}~\cite{Ball:2017otu}.
  This task is beyond the scope of this study.
  Moreover, it has to be noticed that the difference between schemes is due
  to a different treatment of subleading contributions,
  and is thus reduced including higher orders.
  Therefore, when including the all-order resummation of small-$x$ logarithms,
  we expect different schemes to lead to more similar results than in the fixed-order case,
  at least in the small-$x$ region.}
and, therefore, we will also provide results for the FONLL scheme at fixed order.
Since the use of \texttt{APFEL} makes \texttt{xFitter} faster, we migrate to it immediately,
and perform all the following studies within this setup.
Incidentally, the difference between the two schemes also probes an uncertainty
on the PDF extraction, even though it is not related to the parametrization,
but rather to the perturbative uncertainty of the theoretical description.

\begin{table}
\centering
\begin{tabular}{lll}
  Differences in the fit setup & Setup of Sect.~\ref{sec:fitFO}, same as~\cite{Abramowicz:2015mha} & New setup, same as~\cite{Abdolmaleki:2018jln}  \\
  \midrule
  heavy flavour scheme & TR & FONLL \\
  initial scale $\mu_0$ & $1.38$~GeV & 1.6~GeV \\
  charm matching scale $\mu_c$ & $m_c$ & $1.12m_c$ \\
  charm mass $m_c$ & 1.43~GeV & 1.46~GeV
\end{tabular}
\caption{Summary of the differences in the theoretical setup between
  the fit of Sect.~\ref{sec:fitFO} (which is the same of HERAPDF2.0~\cite{Abramowicz:2015mha})
  and the new fits presented in this and in the following sections
  (which is the same of Ref.~\cite{{Abdolmaleki:2018jln}}).}
\label{tab:diff}
\end{table}
The migration from the TR scheme to the FONLL scheme has been already presented in Ref.~\cite{Abdolmaleki:2018jln},
where it was needed for the very same reason (inclusion of small-$x$ resummation in the fit).
It was noted there that changing scheme without acting on any other theoretical parameter
has a mild effect on the PDFs while it leads to a sizeable deterioration of the $\chi^2$.
This result was obtained using the old parametrization Eq.~\eqref{eq:HERApar},
and we found similar results also using our new parametrization Eq.~\eqref{eq:NewPar}.
In order to prepare the code for the inclusion of small-$x$ resummation, however,
another couple of changes were (and are) needed.
One of them is raising the initial scale $\mu_0$ from the HERAPDF2.0 value $\mu_0=1.38$~GeV ($\mu_0^2=1.9$~GeV$^2$)
to $\mu_0=1.6$~GeV, due to a (perturbative) instability of \texttt{HELL} at very small scales.\footnote
{The results provided by \texttt{HELL} become unreliable when $\as\gtrsim0.35$,
  that happens roughly at a scale $\mu\sim1.5$~GeV.
  Note that this instability must not be seen as a limitation of \texttt{HELL},
  but rather as the breakdown of perturbation theory,
  which is obviously more manifest when dealing with all-order quantities.}
As a consequence, since the new initial scale is larger than the
charm mass $m_c=1.43$~GeV, the charm PDF must be generated perturbatively at a
matching scale $\mu_c>\mu_0>m_c$, which then needs to be larger than the default value $\mu_c=m_c$.
The choice adopted in Ref.~\cite{Abdolmaleki:2018jln} and used also here is $\mu_c/m_c=1.12$.
In Ref.~\cite{Abdolmaleki:2018jln} it has also been noted that the use of FONLL
favours a larger value of the charm mass, $m_c=1.46$~GeV, which is then used in that study.
For consistency with Ref.~\cite{Abdolmaleki:2018jln} we also adopt here this value of $m_c$ from now on.
The differences of the fit setups are summarized in Tab.~\ref{tab:diff}.
We stress that since we work at NNLO accuracy, we use the FONLL-C incarnation of this scheme~\cite{Forte:2010ta}.

\begin{table}
\centering
\begin{tabular}{lcc}
  Contribution to $\chi^2$ & Old parametrization~\cite{Abdolmaleki:2018jln} &  New parametrization  \\
  \midrule
  subset NC $e^+$ 920    $\tilde\chi^2/\rm{n.d.p.}$   & $451/377$   & $406/377$   \\
  subset NC $e^+$ 820    $\tilde\chi^2/\rm{n.d.p.}$   & $ 68/ 70$   & $ 74/ 70$   \\
  subset NC $e^+$ 575    $\tilde\chi^2/\rm{n.d.p.}$   & $220/254$   & $222/254$   \\
  subset NC $e^+$ 460    $\tilde\chi^2/\rm{n.d.p.}$   & $218/204$   & $225/204$   \\
  subset NC $e^-$        $\tilde\chi^2/\rm{n.d.p.}$   & $215/159$   & $217/159$   \\
  subset CC $e^+$        $\tilde\chi^2/\rm{n.d.p.}$   & $ 44/ 39$   & $ 37/ 39$   \\
  subset CC $e^-$        $\tilde\chi^2/\rm{n.d.p.}$   & $ 57/ 42$   & $ 50/ 42$   \\
  correlation term + log term     & $100+15$ & $79+2$ \\
  \bf\boldmath Total $\chi^2/\rm{d.o.f.}$  &\boldmath $1388/1131$ &\boldmath $1312/1127$ \\
\end{tabular}
\caption{Same as Tab.~\ref{tab:chi2TR}, but using the FONLL scheme rather than the TR scheme,
  and having raised $\mu_c/m_c=1.12$, $\mu_0=1.6$~GeV and $m_c=1.46$~GeV, namely the setting of Ref.~\cite{Abdolmaleki:2018jln}.}
\label{tab:chi2FONLL}
\end{table}
Within this new setup, we have performed a fit with the old parametrization Eq.~\eqref{eq:HERApar}
(that reproduces the result of Ref.~\cite{Abdolmaleki:2018jln})
and with our new parametrization Eq.~\eqref{eq:NewPar}.
The results in terms of $\chi^2$ are shown in Tab.~\ref{tab:chi2FONLL}.
Comparing with Tab.~\ref{tab:chi2TR} we see that the fit quality
using the FONLL scheme is systematically worse.\footnote
{As commented already in Ref.~\cite{Abdolmaleki:2018jln},
such a difference is mostly due the perturbative contributions included in either schemes,
especially in the computation of the longitudinal structure function.}
We note that the deterioration is worse with the old parametrization,
where the total $\chi^2$ increases by 25 units,
while the new parametrization softens the difference, with an increase of just 11 units.
This is a first indication that the new parametrization is more robust with respect to the HERAPDF2.0 one.
Note that in this scheme the $\chi^2$ reduction when switching from the old parametrization to the new one
is of 76 units, thus larger than in the TR scheme.
The difference at the PDF level between the two fits using the two parametrizations is very similar
to the one found in the TR scheme (Fig.~\ref{fig:TR}), and it is therefore not reported.

\begin{figure}[t]
  \centering
  \includegraphics[width=0.328\textwidth,page=1]{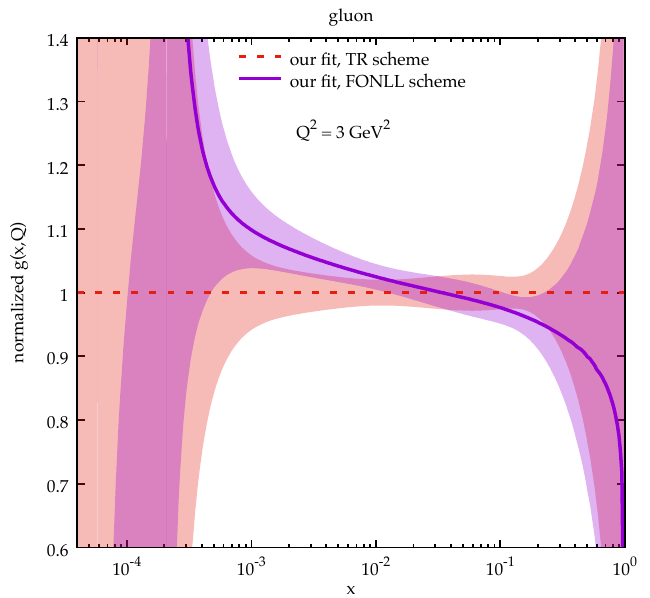}
  \includegraphics[width=0.328\textwidth,page=3]{TR_FONLL_comparison.pdf}
  \includegraphics[width=0.328\textwidth,page=5]{TR_FONLL_comparison.pdf}\\
  \includegraphics[width=0.328\textwidth,page=2]{TR_FONLL_comparison.pdf}
  \includegraphics[width=0.328\textwidth,page=4]{TR_FONLL_comparison.pdf}
  \includegraphics[width=0.328\textwidth,page=6]{TR_FONLL_comparison.pdf}
  \caption{Ratio of some PDFs (same as Fig.~\ref{fig:TR}) obtained using the FONLL scheme
    to the same in the TR scheme, using in both cases our new parametrization.}
  \label{fig:TRFONLL}
\end{figure}
Despite the many differences (Tab.~\ref{tab:diff}), it is interesting
to compare the PDFs obtained in the two setups, in both cases using our new parametrization.
This is done in Fig.~\ref{fig:TRFONLL} in the form of ratios between the two sets at the scale $Q^2=3$~GeV$^2$.
Some differences are manifest, especially for the gluon and the sea quarks.
However, in all cases the 1$\sigma$ bands overlap or are very close to each other, with the single exception
of $\bar d$ $(+\bar s)$ at large $x$, where the absolute PDF is very close to zero (see Fig.~\ref{fig:TR})
and the comparison is therefore not very significant.
Because of the different scheme, this comparison cannot be used to
validate the robustness of the parametrization.
Rather, the similarity of the two PDFs allows us to conclude that the
variations that we are going to perform
are not biased by the choice of using the FONLL scheme rather than the TR scheme.

We thus move to the study of the variations of the fit setting.
First we consider a variation of the fit scale $\mu_0$ up and down.
Specifically we choose to decrease the initial scale to $\mu_0=1.38$~GeV,
which is the same initial scale used in HERAPDF2.0,
and to increase it to $\mu_0=1.84$~GeV,
which is right below the first $Q^2$ bin of HERA data included in the fit
($Q^2=3.5$~GeV$^2$, i.e.\ $\sqrt{Q^2}=1.87$~GeV),
so that it allows us not to cut any data in this variation.
When increasing the scale, we also need to
make sure that the condition $\mu_c>\mu_0$ is satisfied,
so that we can generate the charm PDF perturbatively during DGLAP evolution.
We therefore change $\mu_c/m_c=1.27$, so that $\mu_c=1.85$~GeV,
which is larger than $\mu_0$ but still lower than the value of the first bin with data included in the fit.
In order to disentangle the effect of raising $\mu_c$ (probing a perturbative uncertainty)
and of raising $\mu_0$ (probing a potential parametrization bias),
we also consider an intermediate step when $\mu_c$ is increased but the initial scale is kept fixed at $\mu_0=1.6$~GeV.

\begin{figure}[t]
  \centering
  \includegraphics[width=0.328\textwidth,page=9 ]{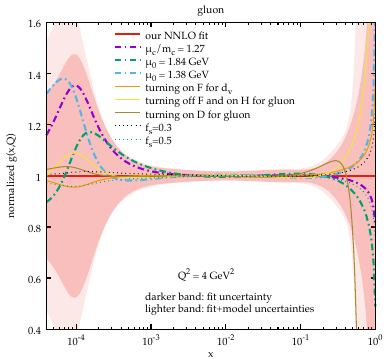}
  \includegraphics[width=0.328\textwidth,page=11]{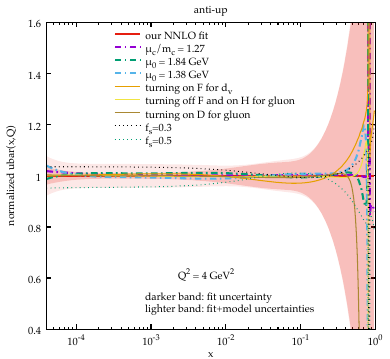}
  \includegraphics[width=0.328\textwidth,page=15]{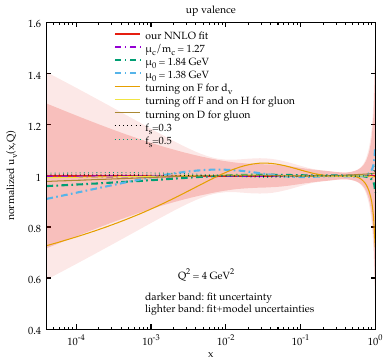}\\
  \includegraphics[width=0.328\textwidth,page=10]{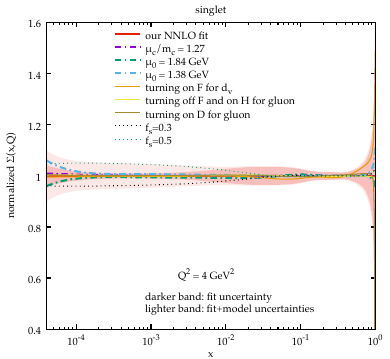}
  \includegraphics[width=0.328\textwidth,page=14]{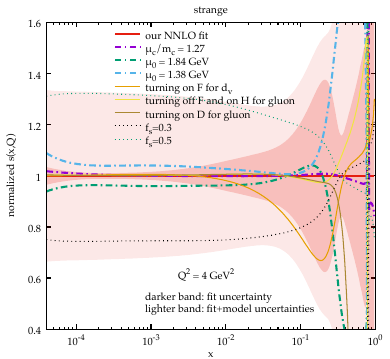}
  \includegraphics[width=0.328\textwidth,page=16]{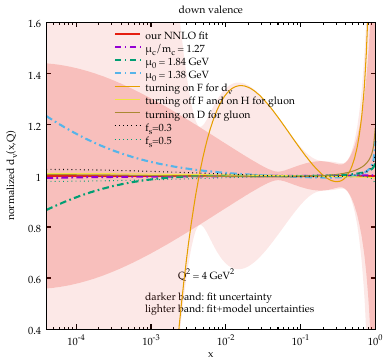}
  \caption{Individual variations of scales and parameters considered. The results are presented as ratios
    with respect to the central fit, at the scale $Q^2=4$~GeV$^2$.
    Differently from Fig.~\ref{fig:TR} and Fig.~\ref{fig:TRFONLL},
    the middle bottom plot shows the strange PDF rather than the combination $\bar d+\bar s$.
    The darker band represents the ``experimental'' (fit) uncertainty,
    while the lighter band is its sum in quadrature with the individual variations.}
  \label{fig:variations}
\end{figure}
The results of these variations are shown in Fig.~\ref{fig:variations},
at the scale $Q^2=4$~GeV$^2$, which is higher than before because it has to be larger than the
largest initial scale considered.
Raising $\mu_c$ (dot-dashed purple) has a mild effect on the quark PDFs,
while it has a stronger impact on the gluon PDFs at small $x$.
This is expected because most of the effect of changing $\mu_c$ is on the charm PDF,
which is in turn mostly determined by the gluon.
For the gluon, raising $\mu_0$ (dot-dashed green) has a small effect compared to
the (dot-dashed purple) curve with the same $\mu_c$,
with the exception of the very small- and very large-$x$ regions.
The effect is similar (though opposite in sign) when lowering $\mu_0$ (dot-dashed cyan).
In both cases the variation is within the ``experimental'' fit uncertainty band (darker red band),
and so it is not very significant.
We have also verified that introducing the cubic logarithmic term proportional to $H_g$ in the parametrization,
that provides additional freedom in the small-$x$ region, does not change the result,
giving us good confidence that the observed initial-scale dependence is not
due to a parametrization bias at small-$x$.

For the quark PDFs, the effect of varying $\mu_0$ at small $x$ is generally mild,
even though some effect is visible.
This is particularly true for the strange PDF.
But this expected, and it highlights indeed a strong bias that we put in our fit:
we fixed the strange PDF to be a fraction of the $\bar d$ PDF at the initial scale, Eq.~\eqref{eq:sdef}.
Because of DGLAP evolution,
at any other scale the strange will not be a fraction (and certainly not the same fraction) of the $\bar d$.
In principle, one should compensate the variation of $\mu_0$ with a variation of Eq.~\eqref{eq:sdef}:
the simplest, though not exact, option would be a variation of $r_s$ (equivalently $f_s$).
Without doing so, the $\mu_0$ variation indirectly
probes the uncertainty on our choice of fixing $r_s$ to a given value.

In fact, variations of the strange ratio $r_s$ (or $f_s$) must be considered in our uncertainty.
Following HERAPDF2.0~\cite{Abramowicz:2015mha}, we consider two variations,
$f_s=0.3$ and $f_s=0.5$, shown in the plots with dotted black and green lines, respectively.
The impact of this variation is negligible on the gluon and the valence PDFs.
However, it has a sizeable impact on all the sea quark PDFs, because of the intimate connection
between the various parameters, see Sect.~\ref{sec:newPar}.
Of course, the largest impact is on the strange PDF itself, acquiring a
very large uncertainty due to this variation.
Even in the total singlet the effect of the variation is well outside the
experimental uncertainty in the region $x<10^{-2}$.
We stress that the $\chi^2$ remains unchanged for these variations,
which is a confirmation that the data considered do not have the power
of constraining the strange PDF.

We now move to the impact of adding (or removing) parameters from our default parametrization Eq.~\eqref{eq:NewPar}.
We have already anticipated in Sect.~\ref{sec:newPar} that we have played with the parameters
and identified three of them ($F_{d_v}$, $D_g$, $H_g$) that give the most significant effects
on the PDFs when they are turned on without affecting the fit quality.
The resulting PDFs are shown in the same Fig.~\ref{fig:variations},
identified by thin solid lines.
The addition of the logarithmic term to the down-valence distribution
has the largest effect. This is accompanied by a reduction of $\chi^2$
by a unit, which makes this parametrization as a potential candidate for the default parametrization.
However, with this extra parameter the $d_v$ distribution becomes negative at small $x\lesssim10^{-3}$,
which is undesired.
For this reason, we decided to keep the simpler parametrization as default,
and activate $F_{d_v}$ for estimating the parametrization uncertainty.
Note that this parameter has an impact also on the up-valence distribution,
and to some extent on the sea distributions (see e.g.\ the strange at medium/large $x$).
The other two parametrization variations are on the gluon PDF.
In one case $D_g$ is activated, allowing more flexibility at large $x$.
Indeed the large-$x$ shape changes substantially, but in a region where
the gluon PDF is very small and largely unconstrained.
In the other case the cubic logarithmic term $H_g$ is activated and the linear logarithmic $F_g$ term
is simultaneously deactivated, as explained in Sect.~\ref{sec:newPar}.
This could in principle make an effect at small $x$, since the flexibility in that region
is obtained through a different functional form.
In practice, the effect is very mild, which is a strong confirmation that our parametrization is robust.

\begin{figure}[t]
  \centering
  \includegraphics[width=0.328\textwidth,page=1]{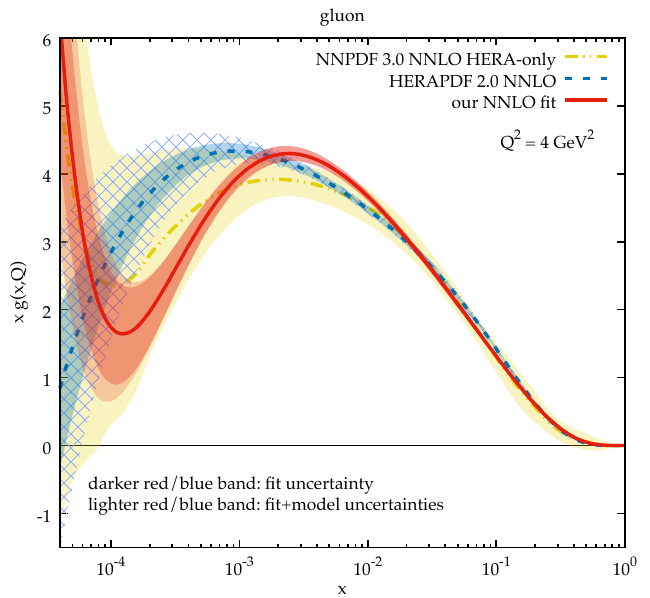}
  \includegraphics[width=0.328\textwidth,page=3]{variations.pdf}
  \includegraphics[width=0.328\textwidth,page=7]{variations.pdf}\\
  \includegraphics[width=0.328\textwidth,page=2]{variations.pdf}
  \includegraphics[width=0.328\textwidth,page=4]{variations.pdf}
  \includegraphics[width=0.328\textwidth,page=8]{variations.pdf}
  \caption{Similar to Fig.~\ref{fig:TR}, but with our new fit in the FONLL scheme, and including both experimental and model uncertainties,
    at the scale $Q^2=4$~GeV$^2$.}
  \label{fig:unc}
\end{figure}
Having all these variations at hand, we can combine them into a (symmetric) uncertainty band.
To do so we sum in quadrature the experimental (fit) uncertainty with each
model uncertainty computed as the difference
between the varied and the central fit,\footnote
{An alternative option would be to consider separately the positive and negative variations,
and construct an asymmetric uncertainty band, as done in Ref.~\cite{Abramowicz:2015mha}.}
with two exceptions.
One is the $f_s$ variation, for which we consider only the $f_s=0.5$ variation,
and we discard the other as it basically gives
the same effect in the opposite direction, so including it would double-count the effect.
The other is the up variation of $\mu_0$, for which we take the difference with respect to the fit with larger $\mu_c$
(so that it measures purely the $\mu_0$ variation), and we discard the uncertainty from $\mu_c$ variation,
since our goal is to construct an uncertainty due to the parametrization choice.
This band is shown as the lighter band of Fig.~\ref{fig:variations}, while the darker one
is just the fit uncertainty.
In Fig.~\ref{fig:unc} we also show the actual PDFs at a the same scale $Q^2=4$~GeV$^2$.
For comparison, we also include the same NNPDF3.0 set considered also in Fig.~\ref{fig:TR},
and the HERAPDF2.0 set. For the latter we also show (with a light blue pattern)
an uncertainty band that includes both the experimental and the model uncertainties,
obtained according to our construction, considering only $\mu_0$, $f_s$ and parameter variations.

We observe that in various regions the inclusion of the variations (lighter red band)
enlarges the fit uncertainty (darker red band) of our PDF set in a substantial way.
For instance, the sea quark distributions at medium/small $x$ have a larger band,
mostly driven by the $f_s$ variations.
The uncertainty on the $d_v$ PDF increases significantly due to the $F_{d_v}$ variation,
so that now the 1$\sigma$ bands of our set and the NNPDF one overlap in all regions of $x$.
The gluon uncertainty band increases visibly only in the small-$x$ region,
mostly due to the initial scale $\mu_0$ variations.
It is interesting to observe that the corresponding uncertainty of HERAPDF2.0 is much larger,
again driven by the initial scale $\mu_0$ variations.
In other cases, like for the down-valence distribution where our uncertainty is large due to the logarithmic term variation,
the HERPDF2.0 uncertainty is smaller than ours.

In conclusion, these studies show that the parametrization that we propose is quite robust
for variations of both the fit scale and of parameters.
The addition of the uncertainty due to these variations provides a result
which appears to be more reliable and robust than what can be obtained with the parametrization Eq.~\eqref{eq:HERApar}.
The main limitation of our result is due to our assumptions on the strange distribution,
which is however a consequence of our dataset not being able to constrain it,
and would be solved by parametrizing the strange PDF independently and fitting a larger dataset with constraining power on it.
Our final PDF set including the full uncertainty
is largely compatible with the NNPDF3.0 set.
These considerations do not exclude the possibility of further improving the functional form
of the PDFs\footnote
{A strong correlation between the $F$ and $G$ parameters has been found,
  which is expected, because the $\log x$ and $\log^2x$ terms are relevant in the same low-$x$ region.
  A way to improve the parametrization is then to find a new definition of the parameters
  such that the correlation is reduced. This would guarantee a better numerical stability
  (even though we stress that we did not encounter problems in the $\chi^2$ minimization procedure),
  without changing the actual functional form.}
and further reduce the parametrization bias,
but they certainly show that our parametrization represents a step forward
with respect to the default \texttt{xFitter} (i.e.\ the HERAPDF2.0) parametrization, Eq.~\eqref{eq:HERApar}.

\section{\boldmath PDF determination with small-$x$ resummation}
\label{sec:fitRes}

We have observed that the $\chi^2$ reduction found here when switching to the new
parametrization is similar to the one obtained in a previous \texttt{xFitter} analysis
through the inclusion of small-$x$ resummation~\cite{Abdolmaleki:2018jln}.
To be precise, using the same FONLL scheme, the $\chi^2$ obtained here at NNLO with the new parametrization (Tab.~\ref{tab:chi2FONLL})
is four points smaller than the one obtained in Ref.~\cite{Abdolmaleki:2018jln}
with the inclusion of small-$x$ resummation using the old parametrization.
This observation is worrying, because it suggests that the improvement found
in Ref.~\cite{Abdolmaleki:2018jln} may not be due to the better theoretical description,
but to the fact that the ``resummed gluon'' is better described through the old parametrization Eq.~\eqref{eq:gluonOldPar}
than it is the ``NNLO gluon''.

This pessimistic interpretation is contradicted by the NNPDF study on small-$x$ resummation~\cite{Ball:2017otu},
where the analysis is based on the unbiased NNPDF parametrization
and the improvement in the description of the HERA data is of similar significance.
However, this does not exclude that at least a part of the improvement found in Ref.~\cite{Abdolmaleki:2018jln}
is due to resummed PDFs having a simpler shape than NNLO PDFs.
Moreover, we have observed in Fig.~\ref{fig:datacomp} that much of the improvement
in the $\chi^2$ when using our new parametrization comes from a better description
of the low-$x$ low-$Q^2$ data which are also responsible for the success of small-$x$ resummation~\cite{Ball:2017otu,Abdolmaleki:2018jln}.
Therefore, the only way to resolve this ambiguity is to perform a PDF fit
with small-$x$ resummation using our new parametrization and compare it with our fixed-order result.

The inclusion of small-$x$ resummation in a PDF fit to HERA data can be obtained using the
\texttt{HELL} code~\cite{Bonvini:2016wki,Bonvini:2017ogt,Bonvini:2018xvt,Bonvini:2018iwt},
that provides the resummed contributions to the DGLAP splitting functions,
the heavy quark matching conditions and the DIS coefficient functions
at the next-to-leading logarithmic accuracy (NLL$x$).\footnote
{To be precise, there are objects that are trivial at LL$x$, e.g.\ all the DIS coefficient functions.
Therefore, NLL$x$ accuracy in these cases represents the first non-vanishing logarithmic contributions,
sometimes referred to as \emph{relative}~LL$x$.}
The current version of \texttt{HELL}, 3.0, differs from the previous version 2.0
(used in the previous resummed fits~\cite{Ball:2017otu,Abdolmaleki:2018jln}) in two respects.
First, it fixes a ``bug'' on the resummation that affected the resummed contributions at NLL$x$
beyond $\Ord(\as^2)$, and is therefore crucial for the matching to fixed-order beyond NNLO.
When resummation is matched up to NNLO (as in all \texttt{HELL} 2.0 studies) the effect of the correction
of the bug is mild (see Ref.~\cite{Bonvini:2018xvt}).
The second difference is due to the introduction of a new default treatment of subleading logarithmic contributions.
One of the basic ingredients used for the resummation (the largest eigenvalue $\gamma_+$ of the singlet anomalous dimension matrix)
was included at an accuracy denoted LL$'$ in \texttt{HELL} 2.0,
and it has been changed to full NLL in \texttt{HELL}~3.0. The difference generated by this change
is formally subleading (i.e., NNLL$x$ on the splitting, matching and coefficient functions),
and so both options (NLL and LL$'$) can be in principle used, and they are indeed both accessible in \texttt{HELL} 3.0.
Based on the behaviour of the $\as$ expansion of the resummed result,
it has been suggested in Refs.~\cite{Bonvini:2018xvt,Bonvini:2018iwt} to use the NLL variant
for default predictions and the LL$'$ variant for assessing the uncertainty from subleading terms.

\begin{table}
\centering
\begin{tabular}{lccc}
  Contribution to $\chi^2$ & \texttt{HELL3.0} (NLL) &  \texttt{HELL3.0} (LL$'$) & \texttt{HELL2.0} (LL$'$)  \\
  \midrule
  subset NC $e^+$ 920    $\tilde\chi^2/\rm{n.d.p.}$   & $402/377$   & $403/377$   & $403/377$ \\
  subset NC $e^+$ 820    $\tilde\chi^2/\rm{n.d.p.}$   & $ 70/ 70$   & $ 69/ 70$   & $ 69/ 70$ \\
  subset NC $e^+$ 575    $\tilde\chi^2/\rm{n.d.p.}$   & $219/254$   & $219/254$   & $218/254$ \\
  subset NC $e^+$ 460    $\tilde\chi^2/\rm{n.d.p.}$   & $223/204$   & $224/204$   & $224/204$ \\
  subset NC $e^-$        $\tilde\chi^2/\rm{n.d.p.}$   & $219/159$   & $220/159$   & $220/159$ \\
  subset CC $e^+$        $\tilde\chi^2/\rm{n.d.p.}$   & $ 38/ 39$   & $ 38/ 39$   & $ 38/ 39$ \\
  subset CC $e^-$        $\tilde\chi^2/\rm{n.d.p.}$   & $ 49/ 42$   & $ 49/ 42$   & $ 49/ 42$ \\
  correlation term + log term                         & $73-7$      & $72-11$     & $72-10$ \\
  \bf\boldmath Total $\chi^2/\rm{d.o.f.}$  &\boldmath $1284/1127$ &\boldmath $1283/1127$ &\boldmath $1283/1127$ \\
\end{tabular}
\caption{Same as Tab.~\ref{tab:chi2FONLL}, for three variants of the resummed NNLO+NLL$x$ fit
  in the FONLL scheme and using our new parametrization.}
\label{tab:chi2res}
\end{table}

For these reasons, in this study we have performed three fits with resummation:
a fit using \texttt{HELL} 2.0 to make contact with the previous studies,
a fit using \texttt{HELL} 3.0 in the LL$'$ variant to verify the impact of the bug correction on the fit,
and finally a fit using \texttt{HELL} 3.0 in the NLL variant that is to be considered as the new default.
The $\chi^2$ contributions to the three fits are presented in Tab.~\ref{tab:chi2res}.
The theoretical setting is the same of Sect.~\ref{sec:fitFONLL}, specifically right column of Tab.~\ref{tab:diff},
which is the same setup of the study of Ref.~\cite{Abdolmaleki:2018jln}.
We note immediately that the three fits are of the same quality, with no significant differences among each other.
In particular, the correction of the \texttt{HELL} 2.0 bug does not affect the fit quality at all.

In all cases, the reduction of $\chi^2$ with respect to the NNLO fit (right column of Tab.~\ref{tab:chi2FONLL})
is significant, with 28-29 units less.
This is much smaller than the 72 units of improvement found in Ref.~\cite{Abdolmaleki:2018jln},
which implies that indeed a significant part of that improvement
was due to the simpler-to-fit shape of the resummed PDFs.
Nevertheless, it is remarkable that after the strong reduction of the $\chi^2$ by 76 units
obtained at fixed-order when switching from the old parametrization to the new one in the FONLL scheme (Tab.~\ref{tab:chi2FONLL})
there is still room for a further reduction of 28 units when turning on small-$x$ resummation.
This is even more remarkable when considering that most of the improvement comes from the same
kinematic region, and it then
reinforces the conclusion that the addition of small-$x$ resummation does improve
the theoretical description of HERA data and is thus very important.

\begin{figure}[t]
  \centering
  \includegraphics[width=0.328\textwidth,page=1]{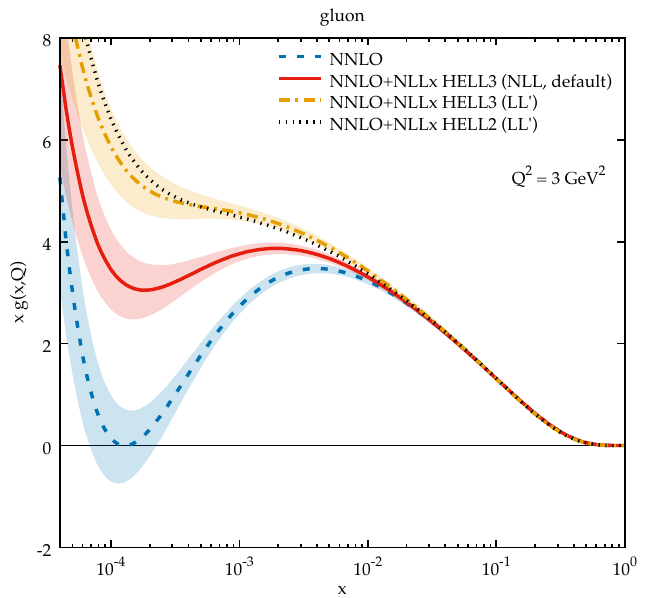}
  \includegraphics[width=0.328\textwidth,page=2]{HELL_comparison.pdf}
  \includegraphics[width=0.328\textwidth,page=5]{HELL_comparison.pdf}\\
  \includegraphics[width=0.328\textwidth,page=7]{HELL_comparison.pdf}
  \includegraphics[width=0.328\textwidth,page=8]{HELL_comparison.pdf}
  \includegraphics[width=0.328\textwidth,page=11]{HELL_comparison.pdf}
  \caption{Comparison of PDFs obtained including small-$x$ resummation from different versions and variants of the \texttt{HELL} code.
    The band represents only the fit uncertainty. The NNLO fit is also shown for reference.}
  \label{fig:HELL}
\end{figure}

We now move to the PDF comparison.
In Fig.~\ref{fig:HELL} we show the gluon, the total singlet and the $u_v$ PDFs
at $Q^2=3$~GeV$^2$ (upper plots),
and the same PDFs in the form of ratios at the electroweak scale $\sqrt{Q^2}=100$~GeV (bottom plots).
Since small-$x$ logarithmic enhancements affect the singlet sector only,
we have focussed on the singlet PDFs, and took a non-singlet one as an example to show
that resummation has basically no effect there.
We observe that at a low scale the correction of the bug
(difference between the \texttt{HELL} 2.0 and \texttt{HELL} 3.0 in the LL$'$ variant)
does not affect the gluon
but it does change the total singlet, even though not dramatically.
After bug correction, the two variants of the resummation (LL$'$ and NLL in \texttt{HELL} 3.0)
give similar results for the quark-singlet at low scale,
while they lead to a rather different gluon.
In particular, the NLL variant predicts a softer gluon at small-$x$,
which is anyway still significantly harder than the NNLO one in the region constrained by the data.
Evolving to the electroweak scale, the effect of the bug is washed out also for the quark singlet,
which is reassuring because it means that the phenomenological applications~\cite{Bonvini:2018ixe,Bonvini:2018iwt,Bertone:2018dse}
performed with resummed PDFs~\cite{Ball:2017otu,Abdolmaleki:2018jln} obtained using \texttt{HELL} 2.0 are reliable.
The NLL variant still leads to a milder effect of resummation, for both the gluon and
the quark-singlet, even though they both remain rather different from their respective NNLO version.
We conclude that even though subleading logarithmic contributions may change the size of the effect of
resummation on the PDFs, the resummed version of the gluon and the quark-singlet PDFs
are always significantly larger at small $x$ than at NNLO.
We finally observe that some differences between NNLO and NNLO+NLL$x$ fits are also present in valence-like PDFs,
even though here the effect is milder and the PDFs are fully compatible within the fit uncertainty.

\begin{figure}[t]
  \centering
  \includegraphics[width=0.65\textwidth,page=1]{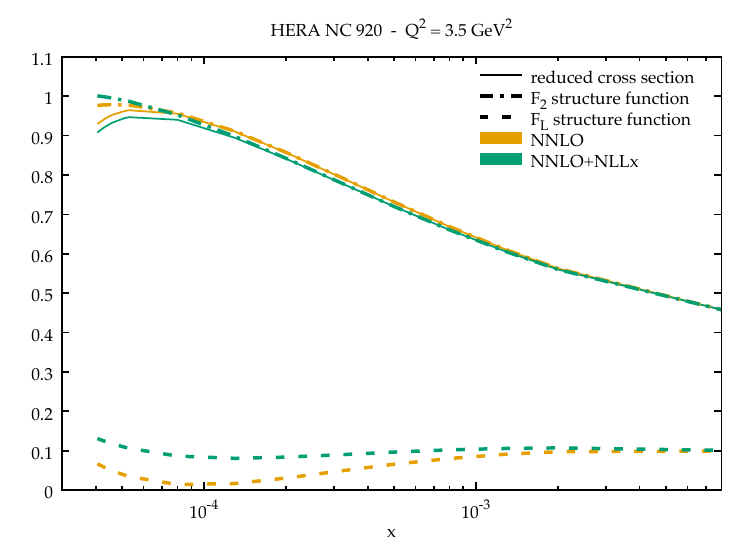}
  \caption{Comparison of the (central) theoretical predictions at NNLO (orange) and at NNLO+NLL$x$ with \texttt{HELL} 3.0 (green)
    for the $Q^2=3.5$~GeV$^2$ neutral-current HERA bin. The plot shows the reduced cross section (thin solid)
    and the structure functions $F_2$ (dot-dashed) and $F_L$ (dashed).}
  \label{fig:F2FL}
\end{figure}

To conclude the discussion, we want to emphasise a difference in the way a good description 
of the low $x$ data is achieved at NNLO and at resummed level with our parametrization.
We have already observed in Sect.~\ref{sec:fitFO} that the NNLO prediction
with our PDFs is able to follow the low-$x$ HERA data in a way that is very similar to
the resummed description reported in Refs.~\cite{Ball:2017otu,Abdolmaleki:2018jln}.
In particular, our parametrization at NNLO is able to reproduce the turnover of the data
at $x\sim10^{-4}$, as shown in Fig.~\ref{fig:datacomp}.
These data are reduced cross sections defined in terms of the structure functions $F_2$ and $F_L$ by
\beq
\sigma_{\rm red}(x,Q^2) = F_2 (x,Q^2) - \frac{y^2}{1 + (1-y)^2} F_L (x,Q^2), \qquad y=\frac{Q^2}{xs},
\eeq
with $\sqrt{s}$ the centre-of-mass energy of the HERA collider.
The success of the resummation in reproducing the turnover was explained
by the harder resummed gluon predicting a larger $F_L$ at small $x$,
that thus gives a larger (negative) contribution to the reduced cross section for $x\lesssim10^{-4}$.
To understand what is the mechanism giving a good description of the same data in our NNLO fit,
we show in Fig.~\ref{fig:F2FL} the theoretical predictions
at NNLO and NNLO+NLL$x$ (using \texttt{HELL} 3.0 in the default NLL variant)
for the $Q^2=3.5$~GeV$^2$ neutral-current HERA bin.
The reduced cross section (thin solid lines) behaves in a very similar way
in both theoretical setups, as they both describe the data with similar accuracy.
However, the behaviour of the two structure functions $F_2$ and $F_L$ is different.
As far as $F_2$ is concerned, the NNLO prediction decreases slightly at small $x\lesssim10^{-4}$,
due to the softer gluon and quark singlet, while at resummed level it rises steadily
due to the harder singlet PDFs.
For the longitudinal structure function the difference is larger: the resummed $F_L$
is quite flat in $x$, and for $x\lesssim10^{-3}$ it is much larger than the NNLO one,
which has a minimum at $x\sim10^{-4}$ where it almost vanishes.
Below this value, both predictions rise,
making the contribution of $F_L$ to the reduced cross section important also at NNLO.
This rise of $F_L$ is due to the shape of the gluon PDF that in our fits
rises below $x\sim10^{-4}$ (see e.g.\ Fig.~\ref{fig:HELL}).
This observation explains why in our NNLO fits the data favour a rising gluon for $x$ below $10^{-4}$,
which is achievable with our flexible parametrization.

\section{Conclusions}
\label{sec:conclusions}

In this work we have proposed a new simple parametrization for the PDFs at the initial scale
that includes a low degree polynomial in $\log x$, Eq.~\eqref{eq:NewParAdd}.
The addition of the polynomial in $\log x$ to the customary polynomial in $x$ (or $\sqrt{x}$) gives
much more flexibility in the low-$x$ region.
We have implemented this parametrization in the \texttt{xFitter} toolkit
and tested it for fitting PDFs from the inclusive HERA I+II dataset,
that counts 1145 datapoints.

We have observed a significant improvement of the fit quality,
with a reduction of the $\chi^2$ by 62 units with respect to the default
\texttt{xFitter} parametrization adopted for instance in the HERAPDF2.0 determination.
This is accomplished using 18 free parameters, to be compared with the 14 parameters used in HERAPDF2.0.
Remarkably, most of the improvement comes from a better description of the gluon PDF,
where the number of free parameters is identical in the two (different) parametrizations.
The quality of our result is also competitive with other mainstream PDF sets.

The PDFs obtained with our new parametrization differ from HERAPDF2.0 in various regions.
The sea-quark PDFs differ mostly at medium $x$, where our parametrization allows for a richer structure.
The up-valence PDF is also quite different for all $x<10^{-1}$.
The major impact is on the gluon PDF, where the shape is also qualitatively different:
at a low scale $Q^2$, our gluon decreases more rapidly for $x\lesssim10^{-2}$
and then it rises below $x\sim 10^{-4}$, while the HERAPDF2.0 gluon keeps decreasing.
Asymptotically, the HERAPDF2.0 gluon grows negative and tends to $-\infty$ as $x\to0$,
while our gluon grows positive and tends to $+\infty$ as $x\to0$.
Incidentally, this behaviour is similar and compatible with the largely unbiased NNPDF determination
from the same data.

We have tested the stability of our parametrization upon variation of the initial scale of the fit
and of the parametrization itself. We have found that our results are very robust,
and generally more stable than HERAPDF2.0.
At the same time, the flexibility of our parametrization also allows for a more
reliable determination of the uncertainties.
In the present study the major limitation is represented by the strange quark distribution,
that is not independently parametrized, introducing a significant bias in our parametrization.
This is a consequence of the fact that the data we use are not sufficient to constrain the strange PDF;
using a larger dataset with constraining power on the strange PDF,
the bias can be simply removed by parametrizing the strange PDF on the same footing as the other fitted PDFs.

Since most of the improvement in the fit comes from a better description of the low-$x$ low-$Q^2$ data,
we have investigated the impact of supplementing the theoretical predictions with the resummation
of small-$x$ logarithms.
Indeed, in previous studies~\cite{Ball:2017otu,Abdolmaleki:2018jln}, the inclusion of small-$x$ resummation
was shown to lead to an improved description of the same data.
We have found that the addition of the resummation further reduces the $\chi^2$ by approximately 30 units,
which is a remarkable achievement given that the $\chi^2$ at fixed order was already rather low with our parametrization.
This confirms the results of the previous studies, namely that the inclusion of small-$x$ resummation
is needed to properly describe low-$x$ low-$Q^2$ data, and puts it on more solid grounds,
as it shows that it was not an accident due to a non-optimal choice of the PDF parametrization.

In the context of the fits with small-$x$ resummation, we have considered three different variants.
One is obtained with a previous version of the resummation \texttt{HELL} code, 2.0,
that was used also in previous studies but contained a bug.
The other two are obtained with the new 3.0 version of \texttt{HELL},
used in a PDF fit for the first time,
and correspond to a bug-fixed version of the previous fit, and to a variant differing by subleading logarithmic contributions.
We have observed that the correction of the bug has overall a very minor effect, mostly concentrated
in the quark singlet at low scale, confirming the validity of the PDF sets obtained
in earlier works~\cite{Ball:2017otu,Abdolmaleki:2018jln}.
The difference induced by subleading contribution is instead substantial, and it changes
the quantitative impact of resummation in the PDF fit.
However, in both cases the effect of the inclusion of small-$x$ resummation is significant,
and leads to PDFs that are not compatible with the fixed-order ones at low $x\lesssim10^{-3}$.

In conclusion, our study shows that a PDF parametrization as simple as our proposal, Eq.~\eqref{eq:NewParAdd},
provides much more flexibility at medium and small $x$ than the standard one adopted in \texttt{xFitter},
reducing the parametrization bias and offering new handles for estimating the PDF uncertainty.
Moreover, in the light of future collider experiments that will probe smaller values of $x$,
our parametrization can also compete with (and be superior to) other parametrizations using
polynomials with non-integer powers of $x$.
We therefore consider our proposed parametrization as a simple, yet important step forward,
and suggest its adoption in the \texttt{xFitter} toolkit.

\acknowledgments
{
We thank Mandy Cooper-Sarkar and Sasha Zenaiev for a critical reading of the manuscript and for various suggestions,
and Sasha Glazov for his support with \texttt{xFitter} and for suggesting to look at the $F_2$ and $F_L$ structure functions.
The work of MB is supported by the Marie Sk\l{}odowska-Curie grant HiPPiE@LHC under the agreement n.~746159.
}

\appendix
\section{Sum rules}
\label{sec:app}

The PDFs must satisfy the so-called quark number sum rules
\begin{align}
  \int_0^1 dx\, u_v(x) &\equiv \int_0^1 dx\[u(x)-\bar u(x)\] = 2 \\ 
  \int_0^1 dx\, d_v(x) &\equiv \int_0^1 dx\[d(x)-\bar d(x)\] = 1
\end{align}
and the momentum sum rule
\beq
  \int_0^1 dx\, x\[ g(x) + \Sigma(x)\] = 1,
\eeq
where $\Sigma(x)$ is the singlet PDF, namely the sum of all the quark and antiquark PDFs.
Introducing the Mellin transform of the PDFs
\beq
\tilde f(N) = \int_0^1 dx\, x^{N-1} f(x),
\eeq
the sum rules can be expressed as
\beq
\tilde u_v(1) = 2, \qquad
\tilde d_v(1) = 1, \qquad
\tilde g(2) + \tilde\Sigma(2) = 1.
\eeq
Computing the Mellin transform of the parametrizations Eq.~\eqref{eq:NewParMult} or Eq.~\eqref{eq:NewParAdd}
is very easy. The generic term behaves as
\beq
f(x) \sim (1-x)^C x^\beta \log^kx,
\eeq
where $k$ is a non-negative integer, and $\beta$ is $B-1$ plus the explicit power of $x$ in the polynomial.
Since $k$ is an integer, we can write the Mellin transform as
\begin{align}
\int_0^1 dx\, x^{N-1} (1-x)^C x^\beta \log^kx
&= \frac{d^k}{dN^k} \int_0^1 dx\, x^{N-1+\beta} (1-x)^C \nonumber\\
&= \frac{d^k}{dN^k} \frac{\Gamma(N+\beta)\Gamma(C+1)}{\Gamma(N+\beta+C+1)}.
\end{align}
Specifically, for $k=1,2,3$ we have ($b=N+\beta$, $c=C+1$)
\begin{align}
  \int_0^1 dx\, x^{b-1} (1-x)^{c-1} \log x 
  &= \frac{\Gamma(b)\Gamma(c)}{\Gamma(b+c)} \Big(\psi_0(b)-\psi_0(b+c)\Big) \\
  \int_0^1 dx\, x^{b-1} (1-x)^{c-1} \log^2 x 
  &= \frac{\Gamma(b)\Gamma(c)}{\Gamma(b+c)} \Big(\[\psi_0(b)-\psi_0(b+c)\]^2 + \psi_1(b)-\psi_1(b+c)\Big) \\
  \int_0^1 dx\, x^{b-1} (1-x)^{c-1} \log^3 x 
  &= \frac{\Gamma(b)\Gamma(c)}{\Gamma(b+c)} \Big(\[\psi_0(b)-\psi_0(b+c)\]^3 + \psi_2(b)-\psi_2(b+c)\nonumber\\
  &\qquad \qquad \qquad + 3 \[\psi_0(b)-\psi_0(b+c)\]\[\psi_1(b)-\psi_1(b+c)\] \Big),
\end{align}
where $\psi_k(z)$ is the polygamma function.

\section{Local minima}
\label{sec:locmin}

During our studies, when fitting the data using fixed-order theory we ended up in a local minimum.
This local minimum is quite far away from the one of the fits reported in the text (that, we hope, is the global one).
The main difference was in the gluon distribution.
In particular, our ``final'' gluon described in the text has a negative $B_g$ parameter,
which is compatible with the theoretical expectations from Regge behaviour.
The local minimum, instead, was characterised by a positive value of $B_g$,
compensated by different values of the parameters of the logarithmic terms,
to give a shape that was qualitatively similar.
Because of the different sign of the $B_g$ parameter, these two minima are very far away from each other in the parameter space.

\begin{table}
\centering
\begin{tabular}{lccc}
  Fitted    & NNLO (FONLL)  &  NNLO (FONLL)   & NNLO+NLL$x$  \\
  parameter & local minimum &  global minimum & \texttt{HELL} 3.0 (NLL)  \\
  \midrule
  $B_g$      & $0.34\pm0.07$   & $-0.55\pm0.03$    & $-0.52\pm0.04$ \\
  $C_g$      & $8.8\pm1.0$     & $4.5\pm0.5$       & $4.5\pm0.5$ \\
  $F_g$      & $0.76\pm0.04$   & $0.230\pm0.003$   & $0.217\pm0.005$ \\
  $G_g$      & $0.22\pm0.02$   & $0.0131\pm0.0004$ & $0.0112\pm0.0005$ \\
  $H_g$      & $0.017\pm0.002$ &   &  \\
  \hline
  $B_{u_v}$   & $0.85\pm0.06$   & $0.83\pm0.06$   & $0.76\pm0.06$ \\
  $C_{u_v}$   & $4.5\pm0.1$     & $4.6\pm0.2$     & $4.6\pm0.1$ \\
  $E_{u_v}$   & $1.7\pm0.8$     & $1.9\pm1.0$     & $2.6\pm1.1$ \\
  $F_{u_v}$   & $0.38\pm0.04$   & $0.37\pm0.05$   & $0.35\pm0.04$ \\
  $G_{u_v}$   & $0.062\pm0.011$ & $0.058\pm0.012$ & $0.049\pm0.010$ \\
  \hline
  $B_{d_v}$   & $1.01\pm0.09$ & $0.98\pm0.10$  & $0.99\pm0.09$ \\
  $C_{d_v}$   & $4.7\pm0.4$   & $4.7\pm0.5$    & $4.7\pm0.5$ \\
  \hline
  $A_{\bar d}$   & $0.070\pm0.008$  & $0.13\pm0.02$   & $0.14\pm0.02$ \\
  $B_{\bar d}$   & $-0.45\pm0.02$   & $-0.34\pm0.02$  & $-0.33\pm0.02$ \\
  $C_{\bar d}$   & $28\pm3$         & $24\pm2$        & $24\pm3$ \\
  $D_{\bar d}$   & $76\pm17$        & $40\pm12$       & $38\pm10$ \\
  $F_{\bar d}$   & $0.084\pm0.001$  & $0.072\pm0.004$ & $0.071\pm0.004$ \\
  \hline
  $C_{\bar u}$   & $11\pm1$   & $11\pm1$  & $11\pm1$ \\
  $D_{\bar u}$   & $33\pm6$   & $20\pm4$  & $18\pm4$ \\
  \midrule
  $\chi^2/\text{d.o.f.}$ & 1314/1126 & 1312/1127 & 1284/1127 
\end{tabular}
\caption{Values of the fitted parameters of our parametrization Eq.~\eqref{eq:NewPar}
  for NNLO fits (first two columns) and a resummed fit (last column).
  The two NNLO results correspond to two separated minima of the $\chi^2$,
  leading to a similar quality of the fit. The values of the $\chi^2$ are also reported.}
\label{tab:params}
\end{table}
To better understand the issue, in Tab.~\ref{tab:params} we have collected the values
of the fitted parameters for three different fits (all using the FONLL scheme):
NNLO theory in the local minimum,
NNLO theory in the global minimum,
and NNLO+NLL$x$ theory (using the default NLL variant of the resummation in \texttt{HELL} 3.0).
We note that the global-minimum NNLO fit and the resummed fit have very similar parameters,
with some differences in the gluon that are not compatible within the fit uncertainty.
Conversely, the fit converged to the local minimum has very significant differences on some of the parameters.
The most striking one is $B_g$, which has opposite sign, but also all the other gluon parameters
differ way more than their uncertainty. Similarly, the sea quark distributions
($\bar d$ and $\bar u$) also present some significant, though less striking, differences.
We stress that the fit that ended up in the local minimum also contained the parameter $H_g$
of the cubic $\log^3x$ term of the gluon, whose presence was crucial to achieve a good description of the data.
Conversely, activating such cubic term in the fit converged to the global minimum
has no impact on the results, as the fit predicts a $H_g$ term that is compatible with zero.\footnote
{Indeed, we initially had $H_g$ as a parameter for the gluon distribution, and we then decided to turn it off
  once we noted that in the global minimum it was not needed anymore.}

The values of the $\chi^2$ are also reported in the table.
We observe that the fit converged to the local minimum has a $\chi^2$ that is very close to the
global minimum, just two units larger.
However, since some parameters are very different in the two fits, the $\chi^2$
becomes much larger when transitioning from one minimum to the other in the parameter space
(in other words, both minima are very deep and surrounded by high ``mountains'').
Therefore, with a standard minimization routine it is highly unlikely that once
the local minimum is found it could converge to the global minimum.
We had to tune the initial values of the parameters by hand so that the minimization
routine started looking directly in a neighborhood of the global minimum
to guarantee that the minimization procedure ends up there.
In this case, the physical expectation $B_g<0$ was crucial to guide us.

\begin{figure}[t]
  \centering
  \includegraphics[width=0.328\textwidth,page=1]{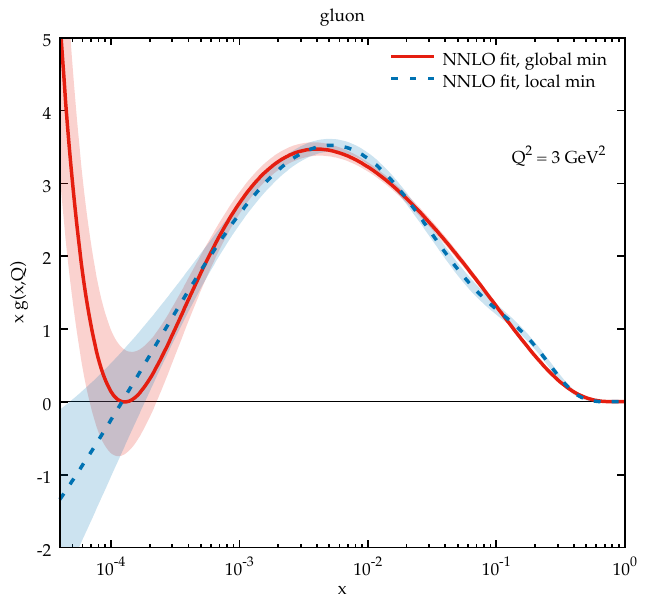}
  \includegraphics[width=0.328\textwidth,page=3]{locmin.pdf}
  \includegraphics[width=0.328\textwidth,page=5]{locmin.pdf}\\
  \includegraphics[width=0.328\textwidth,page=2]{locmin.pdf}
  \includegraphics[width=0.328\textwidth,page=4]{locmin.pdf}
  \includegraphics[width=0.328\textwidth,page=6]{locmin.pdf}
  \caption{Similar to Fig.~\ref{fig:TR}, comparing our NNLO fit using the FONLL scheme (solid red)
    with a variant of the same fit converged to a distant and deep local minimum.}
  \label{fig:locmin}
\end{figure}

Physical considerations apart, the local minimum provides a very decent fit to the data.
For this reason, we believe that it deserves some consideration.
Therefore, we plot in Fig.~\ref{fig:locmin} a comparison of the PDFs converged to the local minimum with
our default ones.
The valence distributions are identical in the two fits, as expected from the comparison of the
corresponding parameters in Tab.~\ref{tab:params}.
The sea quark distributions present small differences in the region around $x\sim10^{-2}$,
and they start to differ significantly below $x\sim 10^{-4}$,
where PDFs from the local minimum bend down, which is a consequence of the larger logarithmic contribution $F_{\bar d}$.
The gluon PDF has a peculiar shape. It follows the global minimum gluon down to $x\sim10^{-4}$,
oscillating around it. Then, below $x\sim10^{-4}$, it keeps decreasing, driven by the cubic logarithmic term,
while the global minimum gluon rises, driven by the asymptotic power term together with the quadratic logarithmic term.
Asymptotically, the local-minimum gluon goes to zero as $x\to0$, because $B_g>0$,
while the global-minimum gluon keeps rising.

We conclude that, over a wide region of $x$, the PDF sets determined by the two minima
are in practice equally good and well compatible.
The differences are mostly concentrated in the small-$x$ region, where indeed there are only few data,
so the PDFs are expected to have somewhat large uncertainties.
One could then be tempted to use the set obtained from the local minimum to compute a more conservative uncertainty.
In principle, if the $\Delta\chi^2$ criterium for computing the uncertainty used $\Delta\chi^2\geq2$,
then this minimum \emph{should} be considered for the uncertainty, even though it would be very difficult for
a numerical routine to discover it while scanning the region around the global minimum.
Moreover, we expect that the uncertainty obtained by scanning around the global minimum
with a larger $\Delta\chi^2$ will not cover the (legitimate!) uncertainty
that would be obtained by also considering the separate local minimum.
Within a hessian fit, there is no simple solution to this problem,\footnote
{A MonteCarlo approach to uncertainty estimation would likely be more suitable for finding local minima
and producing more reliable uncertainties.}
unless the local minimum is found by luck and used for computing the uncertainty by hand.
Without doing so, there is the risk that the PDF uncertainty found from the fit
underestimates the actual uncertainty, that would be another manifestation of a parametrization bias.
We do not have a solution nor we want to propose a recipe,
we just point out that this is an issue to take into account and to treat with care.

\phantomsection
\addcontentsline{toc}{section}{References}
\bibliographystyle{jhep}
\bibliography{references}

\end{document}